\documentclass[12pt]{article}
\pdfoutput=1
\usepackage{color}
\usepackage{graphicx}
\usepackage{amsmath}
\usepackage{slashed}
\usepackage{amssymb}
\usepackage{epstopdf}
\usepackage{caption}
\usepackage{subcaption}

\input xy
\xyoption{all}
\xyoption{web}

\newlength{\abstractwidth}

\renewcommand{\thefootnote}{\fnsymbol{footnote}}
\renewcommand{\thanks}[1]{\footnote{#1}}
\newcommand{\starttext}{
\setcounter{footnote}{0}
\renewcommand{\thefootnote}{\arabic{footnote}}}

\newcommand{\bea}{\begin{eqnarray}}
\newcommand{\eea}{\end{eqnarray}}
\newcommand{\ee}{\end{equation}}
\newcommand{\be}{\begin{equation}}
\newcommand{\ea}{\end{array}}

\newcommand{\half}{\frac{1}{2}}

\DeclareMathOperator{\tr}{tr}

 \newcommand{\reals}{\mathbb{R}}

\begin{document}
\starttext
\setcounter{footnote}{0}

\bigskip

\begin{center}

{\Large \bf Lifshitz black holes in higher spin gravity}
\vskip 0.6in

{ \bf Michael Gutperle, Eliot Hijano and Joshua Samani}

\vskip .2in

{ \sl Department of Physics and Astronomy }\\
{\sl University of California, Los Angeles, CA 90095, USA}\\
\smallskip
{\tt  gutperle@physics.edu;  jsamani@physics.ucla.edu;  eliothijano@physics.ucla.edu}

\vskip .7in

\end{center}

\begin{abstract}
  We study asymptotically Lifshitz solutions to  three dimensional higher spin gravity in  the $SL(3,\reals)\times SL(3,\reals)$ Chern-Simons formulation.  We begin by specifying the most general connections satisfying Lifshitz boundary conditions, and we verify that their algebra of symmetries contains a Lifshitz sub-algebra.  We then exhibit connections that can be viewed as higher spin Lifshitz black holes. We  show that when suitable holonomy conditions are imposed,  these black holes  obey sensible thermodynamics and possess a gauge in which the corresponding metric exhibits a regular horizon.
\end{abstract}

\newpage


\baselineskip=17pt
\setcounter{equation}{0}
\setcounter{footnote}{0}

\newpage

\section{Introduction}
\setcounter{equation}{0}
\label{sec1}

In the past few years there has been a renewed interest in higher spin gravity in
various dimensions
following  the work of Vasiliev and collaborators (see \cite{Vasiliev:2000rn} for
a review). In the
present paper we focus on higher spin theories in three spacetime dimensions.
Gaberdiel and
Gopakumar proposed a duality of the two dimensional  $W_N$ minimal model CFTs to
three
dimensional Vasiliev theory \cite{Gaberdiel:2010pz}. The original proposal has
passed many checks
and some refinements in recent years, see e.g.
\cite{Gaberdiel:2011wb,Gaberdiel:2011zw,Gaberdiel:2012ku,Castro:2011iw,Chang:2011mz,Ammon:2011ua,Perlmutter:2012ds,Hijano:2013fja}.
An
interesting feature of the three dimensional Vasiliev theory
\cite{Prokushkin:1998bq} is that while
it
is
a complicated nonlinear theory coupling an infinite tower of higher spin fields to
scalar matter, if
the
scalars are linearized, the theory can be reformulated in terms of a Chern-Simons
theory with an
infinite dimensional gauge algebra $hs(\lambda)\times hs(\lambda)$
\cite{Blencowe:1988gj,Bergshoeff:1989ns,Pope:1989sr}. The deformation parameter
$\lambda$ is
associated
with
the 't Hooft coupling of the dual CFT \cite{Gaberdiel:2010pz}. The Chern-Simons
theory simplifies if
$\lambda=\pm N$, where $N$ is an integer and the theory reduces to Chern-Simons
theory with
gauge group $SL(N,\reals)\times SL(N,\reals)$ and is purely topological,
corresponding to a theory
of
massless fields of spin $2,3,\cdots,N$. Note that Einstein  gravity with negative cosmological constant is
included by taking
$N=2$
\cite{Witten:1988hc,Achucarro:1987vz}.

 The simplest solutions of the Chern-Simons theory
correspond to $AdS_3$ vacua. The asymptotic symmetry of the $AdS$ vacuum in
$SL(N,\reals)\times
SL(N,\reals)$ higher spin gravity
depends on the embedding of a $SL(2,\reals)$ sub-algebra in  $SL(N,\reals)$. For
the principal
embedding one obtains $W_N$  symmetry \cite{Campoleoni:2010zq,Henneaux:2010xg},
whereas for
non-principal embeddings other higher spin algebras such as $W_N^{(2)}$  can
occur
\cite{Ammon:2011nk,Campoleoni:2011hg}.

The construction of  black holes in $AdS$/CFT is important since (large) black
holes describe the
dual
CFT in thermal equilibrium at finite temperature. The BTZ solution
\cite{Banados:1992wn} of three
dimensional gravity has been a very  important  part of exploring the $AdS$/CFT
correspondence (see
\cite{Kraus:2006wn} for a review). In higher spin theories the definition of what
constitutes a
black
hole is nontrivial since the metric field transforms under higher spin gauge
transformations
\cite{Campoleoni:2010zq} and hence the standard geometric characterization of a black
hole, i.e. the
existence of a horizon is not gauge invariant. In  \cite{Gutperle:2011kf} a new
criterion was
proposed
which uses the holonomy of the Chern-Simons gauge field around the contractable
euclidean time
circle
to characterize a regular black hole.  The holonomy condition has been applied to
various black
holes
in 3 dimensional higher spin theories
\cite{Castro:2011fm,David:2012iu,Chen:2012pc,Chen:2012ba}
 and  it has been checked by comparing bulk and CFT calculations of thermal
 correlation functions
 \cite{Kraus:2012uf,Gaberdiel:2012yb,Gaberdiel:2013jca}, see  \cite{Ammon:2012wc}
 for a review and a
 more extended list of references. Note that there are some puzzles remaining, for
 example there are
 two different proposals for the entropy, namely   the ''holomorphic"
 \cite{Gutperle:2011kf} and
 the
 ''canonical" \cite{Perez:2012cf,Perez:2012cf} one. See
 \cite{Perez:2013xi,Kraus:2013esi,deBoer:2013vca,Ammon:2013hba}
for recent work on the two proposals and their possible relation.

In the Chern-Simons formulation of of higher spin gravity, the $W_N$ extension of
the Virasoro
symmetry of the boundary theory is obtained via the Drinfeld-Sokolov reduction by
specifying
asymptotic boundary fall off conditions for the gauge fields and considering
nontrivial gauge
transformations which respect these boundary conditions. If the boundary conditions
are consistent
then the boundary charges are integrable, finite and conserved and generate the
(extended)
symmetry algebra.

It is a very interesting question whether the higher spin gravity/CFT duality in
three dimensions
can
be generalized to non-AdS backgrounds. In \cite{Gary:2012ms,Afshar:2012nk} a
general recipe and
examples including Lobachevsky ($\reals\times AdS_2$), Lifshitz, Schrodinger and
warped $AdS$
backgrounds were given. More recently the same philosophy was applied to flat
space holography in
\cite{Afshar:2013vka,Gonzalez:2013oaa}.

In the present paper we are interested in a construction and detailed analysis of
higher spin
realizations of asymptotically Lifshitz spacetimes.  Such spacetimes provide
candidates for a
holographic description of field theories with Lifshitz scaling invariance. These
theories  exhibit
an
anisotropic scaling symmetry with respect to space and time ${\vec x}\to \lambda
{\vec{x}}$ and
$t\to
\lambda^z t$, with $z\neq 1$ and are important in various condensed matter systems
(see
\cite{Kachru:2008yh} for references).  In \cite{Kachru:2008yh} a holographic
Lifshitz spacetime
solution  of a gravity theory
coupled to anti-symmetric tensor fields in four dimension was given. Subsequently
Lifshitz space
times
have been ground in many (super)gravity  theories, see e.g.
\cite{Donos:2009en, Balasubramanian:2010uk,Donos:2010tu,Gregory:2010gx}. In holographic theories black
hole or black brane
solutions provide the dual description of field theories at finite temperature
(and chemical
potential
if the black holes are
charged).  For Lifshitz spacetimes the construction of black holes was initiated
in
\cite{Danielsson:2009gi,Bertoldi:2009vn,Mann:2009yx,Balasubramanian:2009rx}, but
most solutions in
the
literature are only known numerically.

In the present paper we focus mainly on the simplest three dimensional higher spin
theory which is
based on $SL(3,\reals)\times SL(3,\reals)$ Chern-Simons theory and corresponds to
gravity coupled to
a
massless spin three field. For simplicity, most explicit calculations are performed
in this theory,
but
we shall also comment on generalizations to  $N>3$ and $hs(\lambda)$.

The structure of the paper is as follows: In section \ref{sec2} we give a brief
review of the
Chern-Simons formulation of higher spin gravity.  In section \ref{sec3} we review
some salient
features
of field theories which enjoy Lifshitz scaling symmetry, and we discuss the
holographic realization
of
such theories. We then review how the Lifshitz spacetime can be obtained as a
solution to
$SL(3,\reals)\times SL(3,\reals)$ Chern-Simons theory, and we demonstrate that the
algebra
generating
Lifshitz isometries can be realized in a higher spin context.

In section  \ref{sec4} we construct black hole solutions with Lifshitz scaling,
focusing on the
simplest case of non-rotating black holes. We discuss the gauge freedom and the
holonomy conditions
as
well as the thermodynamics.
When the holonomy conditions are solved to express the temperature and chemical potential in terms of the extensive parameters there are six different branches.  Only two of  the six  have positive temperature and entropy  and are hence physically sensible.  We consider two additional conditions on the branches, first the
 local thermodynamic stability and second the existence of a radial gauge where the metric exhibits a regular horizon and find that only one branch satisfies all of these conditions.

In section \ref{sec5} we discuss generalizations of our work including the
possibility of
constructing
rotating black hole solutions as well as Lifshitz black holes in $hs(\lambda)$
higher spin theory.

We close with a brief discussion of our results in section \ref{sec6}. For
completeness we
summarize
our conventions for $SL(3,\reals)$ and  $hs(\lambda)$ in an appendix.

\section{Chern-Simons formulation of higher spin gravity}
\setcounter{equation}{0}
\label{sec2}

The Chern-Simons formulation of three dimensional (higher spin) gravity is based
on two copies of the Chern-Simons action at level $k$ and $-k$ and gauge group $SL(N,\reals)\times
SL(N,\reals)$.
\begin{align}\label{chernsimonsa}
  S=S_{CS}[A]-S_{CS}[\bar A]
\end{align}
where
\begin{align}
  S_{CS}[A]= {k\over 4\pi}  \int {\rm tr}\Big( A\wedge dA+{2\over 3} A\wedge A\wedge
  A\Big).
\end{align}
The equations of motion are simply flatness conditions,
\begin{align}
  F=dA+ A\wedge A=0, \qquad \bar F=d\bar A+ \bar A\wedge \bar A=0.
\end{align}
Ordinary gravity is given by the case $N=2$; in the following we will mainly focus
on the case
$N=3$.
This theory was studied in detail in \cite{Campoleoni:2010zq} and it was shown
that the CS theory is
equivalent to $AdS$ gravity coupled to a massless spin three field. The  vielbein
and spin
connection
take values in the $SL(3,\reals)$ Lie algebra and are related to the CS gauge
fields as follows:
\begin{align}\label{vielbein}
  e_\mu ={l\over 2} (A_\mu -\bar A_\mu ), \qquad \omega_\mu=\half (A_\mu +\bar
  A_\mu).
\end{align}
In the following we set the length scale $l$ to one for notational ease.
Using the  expression of the vielbein \eqref{vielbein}  in terms of the
connection,   the metric and
spin 3 field can be expressed as
\begin{align}\label{metform}
  g_{\mu\nu}={1\over 2} \tr(e_\mu e_\nu), \qquad \phi_{\mu\nu \rho} ={1\over 6} \tr(e_{(\mu} e_\nu
e_{\rho)}).
\end{align}
 The gauge transformations act on the Chern-Simons connections as follows
\begin{align}\label{gauget}
  \delta A = d\Lambda + [A, \Lambda], \qquad \delta \bar A = d\bar \Lambda + [A, \bar
\Lambda].
\end{align}
In the construction of asymptotically AdS as well as asymptotically Lifshitz spacetimes, we employ a
special choice of coordinates and choice of gauge. We define a radial coordinate
$\rho$, where the
holographic boundary will be located at $\rho\to \infty$. In addition we define a
timelike
coordinate
$t$ and a space like coordinate $x$, which can be either compact or non-compact
and hence the
boundary
has either the topology of $\reals\times S^1$ or $\reals\times \reals$.  The
``radial gauge" that
we
will use is
constructed by defining $b= \exp( \rho L_0)$ and setting
\begin{align}\label{bigadef}
A_\mu = b^{-1} a_\mu  \,b + b^{-1}\partial_\mu b, \qquad \bar A_\mu = b \,\bar
a_\mu b^{-1} + b
\,\partial_\mu  (b^{-1}).
\end{align}

\section{Lifshitz spacetimes}
\setcounter{equation}{0}
\label{sec3}

Quantum field theories which exhibit a scaling symmetry which is anisotropic with
respect to space and time
\begin{align}
  t\to \lambda^z t, \qquad x\to \lambda x
\end{align}
appear in many condensed matter systems. The dynamical scaling coefficient $z\neq
1$ breaks
relativistic symmetry.  If one augments the symmetry of the theory to include space and time translations, then one obtains a theory that is said to possess Lifshitz symemtry.  Lifshitz symmetry can therefore be encoded as a Lie algebra generated by time
translations $H$,
spatial
translations $P$ and Lifshitz scalings $D$ satisfying the following structure relations:
\begin{align}
  [P,H]&=0 \qquad
  [D,H]=z H \qquad
  [D,P]= P .\label{lifalg}
\end{align}
In two dimensions, conformal symmetry (with $z=1$) implies a conserved, traceless
and symmetric
stress
tensor. For theories with Lifshitz scaling the stress tensor does not have to be
symmetric, since they do not possess boost invariance.  The  stress-energy complex for field theories in 1+1
dimensions with
Lifshitz scaling exponent $z$ contains the following objects:  the energy density
${\cal E}$, the
energy flux ${\cal E}^x$, the momentum density ${\cal P}_x$ and the stress energy
tensor
$\Pi_x^{\;x}$.
These quantities satisfy the following conservation equations (see e.g.
\cite{Ross:2011gu}):
\begin{align}\label{emcoma}
  \partial_t {\cal E}+ \partial_x {\cal E}^x&=0,  \qquad
  \partial_t {\cal P}_x+ \partial_x {\Pi}_x^{\; x}=0.
\end{align}
In addition, the Lifshitz scaling with exponent $z$ implies a modified
tracelessness condition
\begin{align}\label{emcomb}
  z {\cal E}+ {\Pi}_x^{\; x}=0.
\end{align}
The Lifshitz symmetries of a (1+1)-dimensional metric can be realized
holographically with the
following
metric:
\begin{align}\label{lifshitzmet}
  ds^2 = L^2\Big( d\rho^2 -e^{2 z \rho} dt^2 + e^{2\rho} dx^2\Big)
\end{align}
where the Lifshitz scaling transformation corresponds to a translation $\rho \to
\rho+ \ln \lambda$.
This metric is not a solution of Einstein gravity with negative cosmological
constant; one has
to add matter or higher derivative terms to the action to obtain it as a
solution.

One can realize the $z=2$ Lifshitz metric in the $SL(3,\reals)\times SL(3,\reals)$
higher spin
theory \cite{Afshar:2012nk} by choosing the radial gauge as in (\eqref{bigadef}) and
by choosing the
following connections $a=a_\mu \,dx^\mu$ and $\bar a = \bar a_\mu \,dx^\mu$:
\begin{align}\label{lifcon}
  a &= W_2 \,dt+ L_1 \,dx, \qquad   \bar{a} = W_{-2} \,dt + L_{-1} \,dx.
\end{align}
It follows from \eqref{metform} that this connection reproduces the Lifshitz
metric
\eqref{lifshitzmet}
with scaling exponent $z=2$.  Lifshitz spacetimes with critical exponents $z>2$
can be obtained
using
$SL(N,\reals)\times SL(N,\reals)$ Chern-Simons theory with $N>3$.

\subsection{Asymptotically Lifshitz connections} \label{secalif}

Focusing on $N=3$ and $z=2$, we explore Chern-Simons connections that behave
asymptotically like
Lifshitz.  In this section, we use primes to denote derivatives with respect to $x$ and overdots to
denote
derivatives
with respect to
$t$.  Choosing radial gauge as in \eqref{bigadef}, we  look for the most general,
flat connections
with
the property that
\begin{align}
  A - A_\mathrm{Lif} &\sim \mathcal O(1), \qquad\text{as
  $\rho\to\infty$}\label{asympcond}\\
  \bar A - \bar A_\mathrm{Lif} &\sim \mathcal O(1), \qquad\text{as
  $\rho\to\infty$} \label{asympcondbar}
\end{align}
where $A_\mathrm{Lif}$ and $\bar A_\mathrm{Lif}$ are the Lifshitz connections
specified in
\eqref{lifcon}. The most general connections that obey these asymptotics are
obtained by adding
terms
to the Lifshitz connections $a$ in \eqref{lifcon} proprotional to
$W_0,W_{-1},W_{-2}$ and $L_0,
L_{-1}$
(and similarly for $\bar a$).  In particular, we consider the following ansatz:
\begin{align}
  a_t &= W_2 + \ell_{t,0} L_0 + \ell_{t,-1}L_{-1}+
  w_{t,0}W_0+w_{t,-1}W_{-1}+w_{t,-2}W_{-2},
  \label{alant}\\
  a_x &= L_1 + \ell_{x,0} L_0 + \ell_{x,-1}L_{-1}+
  w_{x,0}W_0+w_{x,-1}W_{-1}+w_{x,-2}W_{-2}.
  \label{alanx}
\end{align}
Before applying flatness conditions, we allow all coefficients $\ell_{t,i},
\ell_{x,i}, w_{t,m},
w_{x,m}$ to be arbitrary functions of $t$ and
$x$.  By suitable gauge transformations, we can set
\begin{align}
  w_{x,0} = 0, \qquad w_{x,-1} = 0, \qquad \ell_{x,0} = 0. \label{gfh}
\end{align}
Employing the same notation as used in the higher spin black holes, we denote
\begin{align}
  \ell_{x,-1} =- \mathcal L, \qquad w_{x,-2} = \mathcal W, \label{lwn}
\end{align}
and, after applying the flatness conditions\footnote{See \cite{Henneaux:2013dra}  for discussion of closely related connections and their symmetries.}, we obtain
\begin{align}
  a_t       &= W_2 - 2\mathcal L W_0 + \frac{2}{3}\mathcal L' W_{-1} - 2\mathcal W
  L_{-1} +
  \Big(\mathcal L^2 -\frac{1}{6}\mathcal L''\Big)W_{-2}, \label{cfalt}\\
  a_x       &= L_1-\mathcal LL_{-1} + \mathcal W W_{-2}, \label{cfalx}
\end{align}
along with the following evolution equations for $\mathcal L$ and $\mathcal W$:
\begin{align}
  \dot{\mathcal L} &= 2\mathcal W' \label{evolut1}\\
  \dot{\mathcal W} &= \frac{4}{3} (\mathcal L^2)' - \frac{1}{6}\mathcal L'''. \label{evolut2}
\end{align}
If we follow the same procedure for the barred sector, imposing the condition
\eqref{asympcondbar},
then we find the following asymptotically Lifshitz connections:
\begin{align}
  \bar a_t &= W_{-2}- 2  \bar {\cal L} W_0 -{2\over 3}  \bar {\cal L}' W_{1}+2
  \bar{\cal W}L_1+\Big(
  \bar {\cal L}^2-{1\over 6} \bar  {\cal L}'' \Big)W_{2},\\
  \bar a_x &= L_{-1} - {\cal \bar  L} L_{1}- {\cal \bar W}W_{2}.
\end{align}
where again the flatness conditions produce evolution equations for the barred
quantities
\begin{align}
  \dot {\bar {\mathcal L}} &= -2\bar{ \mathcal W}', \label{evolutbL}\\
  \dot{\bar{ \mathcal W}} &= -\frac{4}{3} ({\bar{\mathcal L}}^2)' +
  \frac{1}{6}{\bar {\mathcal
  L}}''',
  \label{evolutbW}
\end{align}
which can be obtained, up to signs, from \eqref{evolut1} and \eqref{evolut2} by replacing ${\mathcal
L}$ and ${\mathcal
W}$
by $  \bar
{\mathcal L}$ and $\bar {\mathcal W}$.
The signs were chosen so that we can now express the quantities appearing in the
energy-momentum
complex \eqref{emcoma} in terms of the parameters appearing  in the connection as
follows:
 \begin{align}\label{emcomplex}
 {\cal E}  &= {\cal W}+\bar{\cal W},\nonumber  \\
 {\cal P}_x &= {\cal L}-\bar {\cal L},\nonumber  \\
 {\Pi}_x^{\; x} &= -2 {\cal W}-2 \bar{\cal W},\nonumber \\
  {\cal E}^x  &=- \Big({4\over 3}  {\cal L}^2 -{1\over 6} \partial_x^2  {\cal
  L}\Big) +
  \Big({4\over 3}  \bar {\cal L}^2 -{1\over 6} \partial_x^2  \bar {\cal L}\Big).
 \end{align}
 It is straightforward to verify that that evolution equations \eqref{evolut1} and \eqref{evolut2}
imply the equations for the Lifshitz em-complex with $z=2$, given by
\eqref{emcoma} and
\eqref{emcomb}.
\subsection{Realization of Lifshitz symmetries}

We now show that among the gauge transformations that leave the connections
\eqref{cfalt} and
\eqref{cfalx} form-invariant, there exist those that realize the
Lifshitz
algebra as a Poisson algebra of boundary charges.  To begin, recall that for each
gauge parameter
$\Lambda$, the standard definition of the variations of asymptotic symmetry
boundary charges in
Chern-Simons theory is as follows \cite{Campoleoni:2010zq}:
\begin{align}
  \delta Q(\Lambda) = -\frac{k}{2\pi}\int_{-\infty}^\infty dx \tr(\Lambda \delta
  A_x).
  \label{chargedef}
\end{align}
We now show that there exist gauge parameters $\Lambda_H, \Lambda_P, \Lambda_D$
that leave the
asymptotically Lifshitz connections form-invariant.  Moreover, we show that the
variations $\delta
Q(\Lambda_H), \delta Q(\Lambda_P)$ and $\delta Q(\Lambda_D)$ as defined in
\eqref{chargedef} are
integrable and yield charges $ Q(\Lambda_H),  Q(\Lambda_P)$ and $Q(\Lambda_D)$
that realize the
Lifshitz algebra as a Poisson algebra.

As our first step, we determine the most general gauge parameter that results in a
gauge
transformation that leaves the asymptotically Lifshitz connections form-invariant.
TPhe
radial gauge \eqref{bigadef} is
preserved under gauge transformations if and only if the gauge parameter is of the
form
\begin{align}
\Lambda(\rho, t, x) = b^{-1}(\rho)\lambda(t,x) b(\rho).
\end{align}
Given this form, gauge transformations are characterized by the function $\lambda$
and act on the
connections as follows:
\begin{align}
  \delta_\lambda a_\mu = \partial_\mu\lambda + [a_\mu, \lambda].
\end{align}
Now consider a general gauge parameter $\lambda$;
\begin{align}
  \lambda = \sum_{i=-1}^1 \epsilon_i L_i + \sum_{m=-2}^{2}\chi_m W_{m},
  \label{gpp}
\end{align}
where $\epsilon_i = \epsilon_i(t,x)$ and $\chi_m = \chi_m(t,x)$.  Gauge transformations are now explicitly given by
\begin{align}
  \delta_\lambda a_t
  &= -2\delta\mathcal L W_0 + \frac{2}{3}(\delta\mathcal L)'W_{-1} -
  2\delta\mathcal W L_{-1} +
  \left(2\mathcal L\delta\mathcal L - \frac{1}{6}(\delta\mathcal L)''\right)W_{-2},
  \\
  \delta_\lambda a_x
  &= -\delta\mathcal L L_{-1} + \delta\mathcal W W_{-2},
\end{align}
and enforcing form-invariance of the connections allows one to solve for all parameters $\epsilon_i$ and $\chi_i$ in terms of the two parameters
$\epsilon = \epsilon_1$
and $\chi = \chi_2$.
\begin{equation}
\begin{aligned} \label{ezfi}
  \epsilon_0 &= -\epsilon',\\
  \epsilon_{-1} &= -\mathcal L \epsilon + \frac{1}{2}\epsilon'' -2\mathcal W \chi,
\\
  \chi_1 &= -\chi',\\
  \chi_0 &= -2\mathcal L \chi + \frac{1}{2}\chi'', \\
  \chi_{-1} &= \frac{2}{3}\mathcal L'\chi + \frac{5}{3}\mathcal L \chi' -
  \frac{1}{6}\chi''',
  \\
  \chi_{-2} &= \mathcal W \epsilon+ \mathcal L^2\chi - \frac{1}{6}\mathcal L''\chi
  -
  \frac{7}{12}\mathcal L'\chi' - \frac{2}{3}\mathcal L \chi'' +
  \frac{1}{24}\chi''''.
\end{aligned}
\end{equation}
Form-invariance also gives evolution equations for $\epsilon$ and $\chi$
\begin{align}
  \dot\epsilon &= \frac{8}{3}\mathcal L\chi' - \frac{1}{6}\chi''', \\
  \dot\chi &= 2\epsilon',
\end{align}
and it constrains the forms of the variations $\delta\mathcal L$ and
$\delta\mathcal W$
\begin{align}
  \delta\mathcal L &= \epsilon\mathcal L' + 2\epsilon'\mathcal L+ 2\chi\mathcal W'
  + 3\chi'\mathcal
  W
  - \frac{1}{2}\epsilon''',\\
  \delta\mathcal W &= \mathcal W'\epsilon + 3\mathcal W\epsilon' +
  \Big(\frac{4}{3}(\mathcal L^2)' -
  \frac{1}{6}\mathcal L'''\Big)\chi+ \Big(\frac{8}{3}\mathcal L^2 -
  \frac{3}{4}\mathcal
  L''\Big)\chi'
  -
  \frac{5}{4}\mathcal L'\chi'' - \frac{5}{6}\mathcal L\chi'''
  +\frac{1}{24}\chi'''''.
\end{align}
Now that we know the precise form of the most general gauge parameters leaving the connections
form-invariant, we
attempt to identify which of these parameters $\Lambda_H, \Lambda_P$ and $\Lambda_D$ lead
to charges that
satisfy a Lifshitz algebra.  To find these parameters, we first notice that given
the Lifshitz metric \eqref{lifshitzmet}, the Lifshitz algebra is geometrically
realized by the following
killing vectors:
\begin{align}
  \xi_H &= \partial_t, \label{lifkvH}\\
  \xi_P &= \partial_x, \label{lifkvP}\\
  \xi_D &= \partial_\rho - x\partial_x - zt\partial_t.\label{lifkvD}
\end{align}
Explicitly, one easily verifies that
\begin{align}
  [\xi_P, \xi_H] = 0, \qquad [\xi_D,\xi_H] = 2\xi_H, \qquad [\xi_D, \xi_P] =
  \xi_P.
\end{align}
This is precisely the Lifshitz algebra \eqref{lifalg} with $z=2$.  These killing
vectors generate spacetime diffeomorphisms, and there is a standard realization
diffeomorphisms as gauge transformations in Chern-Simons theory via
field-dependent gauge
parameters \cite{Banados:1994tn}
\begin{align}
  \Lambda = -\xi^\mu A_\mu.
\end{align}
For the asymptotically Lifshitz connections of section \ref{secalif}, we expect
that there exists a realization of the Lifshitz algebra, but it is not immediately
obvious which gauge parameters one should pick that yield charges satisfying this
algebra.  However, motivated by the method of generating diffeomorphisms via gauge
transformations, we try the following:
\begin{align}
  \Lambda_H &= -(\xi_H)^\mu A_\mu = b^{-1}(-a_t)b, \\
  \Lambda_P &= -(\xi_P)^\mu A_\mu = b^{-1}(-a_x)b, \\
  \Lambda_D &= -(\xi_D)^\mu A_\mu = b^{-1}(-L_0 + xa_x + 2ta_t)b.
\end{align}
These gauge parameters leave the
asymptotically Lifshitz connections form-invariant because one can show that there
exists choices of
the parameters $\epsilon$ and $\chi$ that lead to these gauge parameters.  To see
this explicitly,
notice that given $\epsilon(t,x)$
and $\chi(t,x)$, if we let $\widehat\lambda(\epsilon(t,x), \chi(t,x))$ denote the
gauge parameter
$\lambda(t,x)$ of \eqref{gpp} obtained after all $\epsilon_i$ and $\chi_m$ have
been
substituted
for their expressions in terms of $\epsilon$ and $\chi$ in
\eqref{ezfi}, hen we have
\begin{align}
  \Lambda_H &= b^{-1} \widehat\lambda(0,-1)b, \\
  \Lambda_P &= b^{-1} \widehat\lambda(-1, 0) b, \\
  \Lambda_D &= b^{-1} \widehat\lambda(x, 2t) b.
\end{align}
We now have candidates for gauge parameters from which to construct charges that
satisfy the Lifshitz
algebra.  Using the definition \eqref{chargedef}, we find that the expressions for
the variations of
the charges corresponding to these gauge parameters are integrable and give
\begin{align}
  Q(\Lambda_H) &= \frac{2k}{\pi}\int_{-\infty}^\infty dx \mathcal W, \\
  Q(\Lambda_P) &= \frac{2k}{\pi}\int_{-\infty}^\infty dx \mathcal L, \\
  Q(\Lambda_D) &= -\frac{2k}{\pi}\int_{-\infty}^\infty dx (2t\mathcal W +x\mathcal
  L).
\end{align}
To determine the Poisson algebra of these charges, we recall that for any two
gauge parameters
$\Lambda$ and $\Gamma$, one has \cite{Campoleoni:2010zq,Banados:1994tn}
\begin{align}
  \{Q(\Lambda), Q(\Gamma)\} = \delta_\Lambda Q(\Gamma).
\end{align}
We assume
that
the fields $\mathcal L$ and $\mathcal W$ vanish sufficiently rapidly as
$x\to\pm\infty$ to ensure
that
any boundary terms encountered in computing the gauge-variations of the charges
vanish.  After some tedious but straightforward calculation, we find that
\begin{align}
  \{Q(\Lambda_H), Q(\Lambda_P)\} &= 0,  \\
  \{Q(\Lambda_D),Q(\Lambda_H) \} &= 2 Q(\Lambda_H), \\
  \{Q(\Lambda_D),Q(\Lambda_P)\} &= Q(\Lambda_P).
\end{align}
This is precisely the Lifshitz algebra \eqref{lifalg}. In two dimensions we
expect
that the Lifshitz algebra will be
extended to an infinite-dimensional algebra, in analogy with the extension of
global conformal
symmetry
to a Virasoro algebra. A proposal for an infinite-dimensional extension of the
Lifshitz symmetry was
made in \cite{Compere:2009qm} and can be investigated using the Chern-Simons
formulation.

\section{Non-rotating Lifshitz black hole}
\setcounter{equation}{0}
\label{sec4}

The most general solutions of the Chern-Simons theory have  connections $A$ and
$\bar A$ which are
independent.   We relate the barred and unbarred charges by setting
\begin{align}\label{conrel}
  \bar a_x = -a_x^T, \qquad  \bar a_t=a_t^T,
\end{align}
 leaving the solutions to be
characterized by only by
the
unbarred connection $a_\mu$.  Consequently, the expression for the
metric \eqref{metform} is diagonal, i.e. the $g_{tx}$ component of the metric vanishes.

\subsection{Most general non-rotating black hole solutions}

Restricting ourselves to $SL(3, \reals)\times SL(3, \reals)$ Chern-Simons, we
start with a
generalization of the ansatz \eqref{alant}, \eqref{alanx} in which we allow for
source terms as
coefficients of the generators $W_2$ and $L_1$ in the temporal components of the
connections.  This
changes the asymptotics, but as we will see presently, this extra freedom will
allow us to interpret
the resulting solutions as finite energy excitations above the asymptotic
Lifshitz vacuum. We also
restrict our attention to coordinate-independent connection coefficients.  Our
general ansatz for the
unbarred sector is
\begin{align}
  a_t &=  \ell_{t,1}   L_1 + w_{t,2}W_2 + \ell_{t,0} L_0 +
  \ell_{t,-1}L_{-1}+w_{t,1}W_1+
  w_{t,0}W_0+w_{t,-1}W_{-1}+w_{t,-2}W_{-2}  \label{bhalant},\\
  a_x &= L_1 + \ell_{x,0} L_0 + \ell_{x,-1}L_{-1}+
  w_{x,0}W_0+w_{x,-1}W_{-1}+w_{x,-2}W_{-2}.
  \label{bhalanx}
\end{align}
 Notice that the ansatz \eqref{alant}, \eqref{alanx} of the last section is a
 special case of this
 ansatz obtained by setting $\ell_{t,1} = 0$ and $w_{t,2} =1 $.  In order for this
 ansatz to be a
 solution of our theory we need to impose the flatness conditions which constrain the connections;
\begin{equation}
\begin{aligned}\label{flat}
  w_{t,1} &= 0 ,\\
  \ell_{x,0} &= 0 ,\\
  w_{t,-1} &=\ell_{t,1} w_{x,-1}  ,\\
  \ell_{t,0} &=-w_{t,2} w_{x,-1} ,\\
  \ell_{t,-1} &=\ell_{t,1} \ell_{x,-1}-2 w_{t,2} w_{x,-2} ,\\
  w_{t,0} &=\ell_{t,1} w_{x,0}+2 \ell_{x,-1} w_{t,2} ,\\
  w_{t,-2} &= \ell_{x,-1}^2
  w_{t,2}+\ell_{t,1}w_{x,-2}+w_{x,0}w_{t,2}w_{x,-2}-\frac{1}{4}w_{2,t}w_{x,-1}^2
  .
\end{aligned}
\end{equation}
These conditions seems to indicate that a flat solution is specified by parameters $\ell_{t,1}, \ell_{x,-1}$ and $w_{t,2},w_{x,0}, w_{x,-1},w_{x,-2}$. However we have not fixed all the gauge
freedom, and some of these
parameters are gauge artifacts. In order to see which of these parameters
are the
charges and sources
of the theory and wich of them can be gauged away, it suffices to look at the only
gauge invariant
quantities of the theory: the holonomies. A quick inspection of the holonomies
around the thermal
and
angular cycles shows that the following quantities distinguish different solutions
\begin{equation}
\begin{aligned}\label{gaugeinva}
\mu_2 &=w_{t,2} , \\
\mu_1 &=\ell_{t,1}+\frac{1}{3}w_{x,0}w_{t,2} ,\\
\cal L &=-\ell_{x,-1}+\frac{1}{12}w_{x,0}^2 ,\\
\cal W &=w_{x,-2}+\frac{1}{54}\left(18\ell_{x,-1}w_{x,0}-w_{x,0}^3\right) .
\end{aligned}
\end{equation}
Under these identifications we will interpret $\mu_1,\mu_2$ and $4{\cal L}, -4{\cal
W}$ as sources
and
their conjugate charges.  We will expand on this interpretation in section \ref{I1I2}.
Finally, to obtain
a
generic solution for the barred sector, we take $\bar A = -A^T$ replacing
$\mu_i$ by $\bar
\mu_i$
and $\cal L$, $\cal W$ by $\cal \bar L$ and $\cal \bar W$. Limiting out attention
to non
non-rotating solutions implies setting $\bar \mu_i=-\mu_i$, $\cal \bar
L = \mathcal L$ and $\cal \bar \cal W$.

Note that for a non-vanishing source $\mu_1$, the connection \eqref{bhalant} has a
nonzero $L_1$
component and does not satisfy the criterion for an asymptotically Lifshitz
connection
\eqref{asympcond}. This indicates that the source $\mu_1$ deforms the Lifshitz
vacuum just as in the
case of the higher spin CFTs.

\subsection{Holonomy conditions}

In the context of Chern-Simons higher spin theories, black hole solutions need to
satisfy certain
holonomy conditions and should have a thermodynamical interpretation \cite{Gutperle:2011kf,Castro:2011fm}.
In particular, the requirement of a smooth Euclidean geometry implies that the
thermal holonomy of
the
Chern-Simons connection is trivial;
\begin{align}
  \mathcal P \exp\left(\oint_t dt A_t\right) = \bold{1}, \label{thol}
\end{align}
where $\bold{1}$ is the $SL(3,\reals)$ identity, and the thermal cycle is from
$t=0$ to $t=2\pi i$.
This condition can be recast in more than one equivalent way.  Diagonalizing
$a_t$, and noting that $a_t$ is constant, we find that the condition of a trivial
thermal holonomy is equivalent to the following condition on the eigenvalues
$\lambda_1$, $\lambda_2$, and $\lambda_3$ of $a_t$;
\begin{align}
  e^{2\pi i\lambda_1} =  e^{2\pi i\lambda_2} =  e^{2\pi i\lambda_3} = 1.
  \label{evalhol}
\end{align}
This means that each eigenvalue of $a_t$ must be an integer.  Since $A_t$ is an
element
of $\mathfrak{sl}(2,\reals)$, it must be traceless, and this gives a second
requirement on the
eigenvalues; they must sum to zero.
\begin{align}
  \lambda_1+\lambda_2+\lambda_3 = 0. \label{evaltr}
\end{align}
The simplest nontrivial solution is then
$(\lambda_1,\lambda_2,\lambda_3)=(0,1,-1)$. This solution
contains the famous BTZ black hole and its higher spin generalizations studied in \cite{Gutperle:2011kf}.

In order to find black hole solutions one demands that the connections
\eqref{bhalant} and
\eqref{bhalanx} obey \eqref{evalhol} and
\eqref{evaltr}. These conditions can be cast in a computationally convenient
light.  Employing the
Cayley-Hamilton theorem, we note that every 3-by-3 complex matrix $X$ satisfies
its own
characteristic
polynomial.  This means that there exist complex numbers $\Theta_0, \Theta_1,
\Theta_2$ for which
\begin{align}
  X^3 = \Theta_0 I+ \Theta_1 X+\Theta_2 X^2.
\end{align}
In particular, this allows one to compute any integer power of $X$ knowing only
the coefficients of
the
characteristic polynomial, and therefore allows for evaluation of the matrix
exponential of $X$ in
terms of these coefficients.  In the special case that $X$ is traceless, which is
the case
for
the argument of the exponential in the thermal holonomy, there are simple
expressions for the
coefficients of the characteristic polynomial, which therefore serve to determine
the thermal
holonomy
completely;
\begin{align}
  \Theta_0 = \det(X), \qquad \Theta_1 = \frac{1}{2}\tr(X^2), \qquad\Theta_2 = 0.
\end{align}
Applying this to the triviality condition \eqref{thol}, we find that the
eigenvalues of $a_t$ are
related to the characteristic polynomial coefficients;
\begin{align}
  \Theta_0 =(2\pi i)^3\lambda_1\lambda_2\lambda_3, \qquad \Theta_1
  =-2\pi^2(\lambda_1^2+\lambda_2^2+\lambda_3^2).
\end{align}
In the case of, for example, the BTZ black hole, with
$(\lambda_1,\lambda_2,\lambda_3)=(0,1,-1)$
one
obtains
\begin{align}
  \Theta_0 = 0, \qquad \Theta_1 = -4\pi^2, \qquad \Theta_2 = 0. \label{btzcay}
\end{align}
In the context of finding a higher spin Lifshitz black hole solutions, we see no
compelling reason
to
choose the BTZ holonomy conditions over others, but we do so anyway because they
are simple and
non-trivial.  In principle, however, any conditions on the eigenvalues $\lambda_j$
satisfying
\eqref{evalhol} and \eqref{evaltr} should give rise to independent solutions.
Applying the conditions \eqref{btzcay} to our solution, we obtain the following
holonomy conditions:
\begin{align}\label{holconds}
  0 &= 3\mathcal{L}\mu _{1}^{2}+9\mathcal{W}\mu _{1}\mu _{2}+4\mathcal{L}^{2}\mu
_{2}^{2}-\frac{3}{4} ,\\
  0 &=  108\mathcal{W}^{2}\mu _{2}^{3}+8\mathcal{L}^{2}\mu _{2}\left( 9\mu
  _{1}^{2}-4%
\mathcal{L}\mu _{2}^{2}\right)+27\mathcal{W}\left( \mu _{1}^{3}+4\mathcal{L}%
\mu _{1}\mu _{2}^{2}\right). \label{holcondsb}
\end{align}
These two equations can be used to solve for any two of ${\cal L},{\cal W},\mu_1,\mu_2$ in terms of the remaining two. In the next section we shall argue that thermodynamically  ${\cal L}$ and ${\cal W}$ are charges and $\mu_1,\mu_2$ are the conjugate potentials.

\subsection{Action and entropy}\label{I1I2}

Since the black holes we are studying are gravitational solutions, we need to
check that the Chern-Simons theory provides a correct variational principle.  Let
$I_0$ denote the euclidean Chern-Simons action.  The on-shell, euclidean action
$I_0^\mathrm{os}$, namely the action in which the equations of motion have been used, is given by a boundary term
\begin{align}
  I_0^\mathrm{os} &= -{k \over 4 \pi}  \int d\phi\, dt \,\tr(a_t a_x)
  +\frac{k}{4\pi}\int d\phi \,dt \,\tr(\bar
 a_t
 \bar a_x),
\end{align}
and evaluating the action on our non-rotating connections gives \cite{Banados:2012ue};
\begin{align}
  I_0^\mathrm{os} = -4 k \big( 2 {\cal L} \mu_1 + 3 {\cal W} \mu_2\big).
\end{align}
However $I_0^\mathrm{os}$ does not obey  a thermodynamically sensible variational
principle because the on-shell variation of $I_0$ is
 \begin{equation}
  (\delta I_0)^\mathrm{os} = 8k \big( {\cal L}\delta \mu_1 + {\cal W} \delta
  \mu_2\big)+\delta(4k\mu_2 \cal
  W).
 \end{equation}
The third term spoils the identification of $\mu_1,\mu_2$ with sources having
conjugate charges ${\mathcal L}$ and ${\mathcal W}$.  As discussed in  \cite{Banados:2012ue}, in the context of the higher spin black
holes, it is possible to obtain a canonical action $I_1$ that is
thermodynamically sensible by adding a boundary term to $I_0$. When we evaluate
$I_1$ on our non-rotating solutions, we obtain
\begin{align}\label{I1}
 I_1^\mathrm{os}&=-8k(\mu_1 {\cal L}+2\mu_2 {\cal W}),
\end{align}
and it has the corresponding on-shell variaton
\begin{align}\label{delI1}
  (\delta I_1)^\mathrm{os}=8k({\cal L}\delta \mu_1+{\cal W}\delta \mu_1).
\end{align}
This relation follows directly from the holonomy conditions \eqref{holconds} and \eqref{holcondsb}. Taking
derivatives of the
conditions with respect to the sources, one can show that
\begin{align}
  \frac{\partial\mathcal L}{\partial\mu_1}&= -{6\mu_1 {\cal L}-18 \mu_2 {\cal W}
  \over 3 \mu_1^2 - 16 \mu_2^2 {\cal
L}}
,
&\frac{\partial\mathcal W}{\partial\mu_2}&= -{ 8 {\cal L}( \mu_1 {\cal L}- 3 \mu_2
{\cal W})  \over 3 \mu_1^2 - 16
\mu_2^2 {\cal L}} , \label{partlawe1}\\
\frac{\partial\mathcal L}{\partial \mu_2}&= {16\mu_2 {\cal L}^2-9 \mu_1 {\cal W}
\over 3 \mu_1^2 - 16 \mu_2^2 {\cal
L}},
&\frac{\partial\mathcal W}{\partial\mu_1}&= {16\mu_2 {\cal L}^2-9 \mu_1 {\cal W}
\over 3 \mu_1^2 - 16 \mu_2^2 {\cal
L}}. \label{partlawe2}
\end{align}
Using these expression one can easily show that  that  $\mu_1,\mu_2$ are conjugate to ${\mathcal L}$ and ${\mathcal
W}$ respectively;
\begin{equation}
 {\partial I_1^\mathrm{os} \over \partial \mu_1} = 8 k {\cal L}, \quad \quad
 {\partial I_1^\mathrm{os}
 \over
 \partial \mu_2} = 8 k {\cal W}.
 \end{equation}
The following integrability relation follows immediately from the equality of mixed partial
derivatives:
\begin{align}
  \frac{\partial\mathcal W}{\partial\mu_1} = \frac{\partial\mathcal
  L}{\partial\mu_2}.
\end{align}
The entropy $S$ is naturally a function of the charges ${\mathcal L},{\mathcal
W}$. It can can be obtained
 by performing a Legendre transform  of $I_1^\mathrm{os}(\mu_1,\mu_2)$ with
respect to the conjugate
variables ${\mathcal L}$ and  ${\mathcal W}$.
\begin{align}
 S({\mathcal L},{\mathcal
W}) &=  {\partial I_1^\mathrm{os} \over \partial \mu_1} \mu_1+  {\partial
 I_1^\mathrm{os} \over \partial
 \mu_2} \mu_2- I_1^\mathrm{os}\nonumber \\
 &= 8k\big( 2 \mu_1 {\cal L}+ 3 \mu_2 {\cal W}\big).\label{entone}
\end{align}
Moreover, using the holonomy conditions one can easily verify that the following inverse
thermodynamic
relations
are:
\begin{equation}\label{entLW}
 {\partial S \over \partial(8 k {\cal L})} = \mu_1 , \quad \quad  {\partial S
 \over \partial(8 k
 {\cal
 W} )}= \mu_2.
\end{equation}

\subsection{Temperature and grand potential}

Recall that for any thermodynamic system, the grand potential is defined as
follows in terms of the thermal partition function:
\begin{align}
  \Phi = -\frac{1}{\beta} \ln Z.
\end{align}
Using the saddle point approximation, we identify the on-shell, Euclidean
Chern-Simons action with the log of the partition function, so we obtain
\begin{align}
  \Phi = \frac{1}{\beta} I_1^\mathrm{os}. \label{phios}
\end{align}
The thermodynamic potential $\Phi$ is associated with the grand canonical ensemble and has as natural variables the temperature $T$ and the chemical potential $\alpha$.  These can be related to $\mu_1,\mu_2$ as follows.

In euclidean signature we have chosen the periodicity of the
euclidean time circle to be $1$.  A different euclidean periodicity $\beta$  is equivalent to keeping the
periodicity equal to $1$ and rescaling $A_t$ by a factor of $\beta$. This leads us to
re-express the
potentials
$\mu_1,\mu_2$ in terms of $\beta$ (or the temperature $T$) and a higher spin
potential $\alpha$.
\begin{align}\label{talre}
\mu_1 = \beta \alpha ={1\over T} \alpha,\quad \quad \mu_2 =\beta={1\over T}.
\end{align}
This prescription also ensures that after the rescaling, the connections have
Lifshitz asymptotics.

In thermodynamics it is a well known fact that
the grand canonical  potential has the following differential
\begin{align}
d\Phi= -S dT - Q d\alpha.
\end{align}
It follows that the entropy $S$ and charge $Q$ can be computed as appropriate partial derivatives of the grand potential;
\begin{align}
  \left. {\partial \Phi\over \partial T}\right|_\alpha=- S , \quad \quad
  \left.{\partial
  \Phi\over\partial  \alpha }\right|_T= -Q .
\end{align}
For the Lifhsitz black hole, the grand potential  $\Phi$ is related to the on shell action $I_1^{\mathrm{os}}$ via  \eqref{phios} which is turn is given  by \eqref{I1}, giving
\begin{align}\label{grandpotential}
\Phi = - 8 k  \Big( \alpha {\cal L} +2 {\cal W}\Big).
\end{align}
Using the holonomy conditions \eqref{holconds} and \eqref{holcondsb} to eliminate derivatives with respect to $\alpha,T$  one can calculate the
entropy
\begin{align}\label{entropythermo}
S= -\left. {\partial \Phi\over \partial T}\right|_\alpha= {1\over T} 8k \Big( 2
\alpha {\cal L} + 3
{\cal W}\Big).
\end{align}
Note that the entropy agrees with \eqref{entone}.
The charge conjugate to the potential $\alpha$  is given by
\begin{align}
Q= -\left.{\partial \Phi\over\partial  \alpha }\right|_T=-8k {\cal L}.
\end{align}
We can use the thermodynamic relation between grand potential and the internal energy (which we can identify with the mass of the black hole) to obtain a formula for the energy $E$\footnote{Notice that it follows from \eqref{efromphi} that the Gibbs-Duhem relation  $E=TS+ \alpha Q$  doesn't hold for our black hole solution, since   it would imply $\Phi=0$. This is a common feature of black hole thermodynamics which has been noticed at various points in the literature (see e.g. \cite{Wright:1980a}\cite{Gibbons:2004ai}).};
\begin{align}
E&= \Phi+ TS+ \alpha Q =8 k {\cal W}. \label{efromphi}
\end{align}
Note that this result agrees with the identification of $\mathcal W$ with the energy in the holographic Lifshitz em-complex given in \eqref{emcomplex}.
We can perform one last consistency check by  solving the holonomy conditions with $S$ and $Q$ as independent variables, it is straightforward to verify that the First Law of thermodynamics is indeed satisfied;
\begin{align}\label{firstlaw}
  dE = TdS + \alpha dQ.
\end{align}

\subsection{Branches}\label{sec:branches}
After clarifying the thermodynamical interpretation of the parameters in the
connection, we are ready to look for black hole solutions to the holonomy conditions \eqref{holconds} and \eqref{holcondsb}.  In this section we will express the  intensive parameters $T$ and $\alpha$  in terms of the extensive parameters ${\mathcal L}$ and ${\mathcal W}$. Note that due to the nonlinear nature of the holonomy conditions, there will be multiple branches which can be interpreted as different phases of the theory.

In order
to
simplify the calculation it proves useful to replace $\cal W$ by a parameter $\theta$ which is given
by
\begin{align}
  {\cal W}=\sqrt{\frac{16{\cal L}^3}{27}}\sin\theta .
\end{align}
We utilize \eqref{entLW} to replace $\mu_1,\mu_2$ in
The first holonomy condition \eqref{holconds}  by derivatives of the entropy with respect to ${\mathcal L}$ and $\theta$.
\begin{align}
  64k^2{\cal L}=9\left(\frac{\partial S}{\partial\theta}\right)^2+4{\cal L}^2\left(\frac{\partial S}{\partial\mathcal L}\right)^2.
\end{align}
This partial differential equation for $S$ is solved by the following family of solutions parametrized by a constant
$C$.
\begin{align}
  S({\cal L},\theta)=8k\sqrt{\cal L}\cos\left(\frac{\theta}{3}+C\right) .
\end{align}
\begin{figure}
  \centering
  \begin{tabular}{ccc}
  \includegraphics[width=50mm]{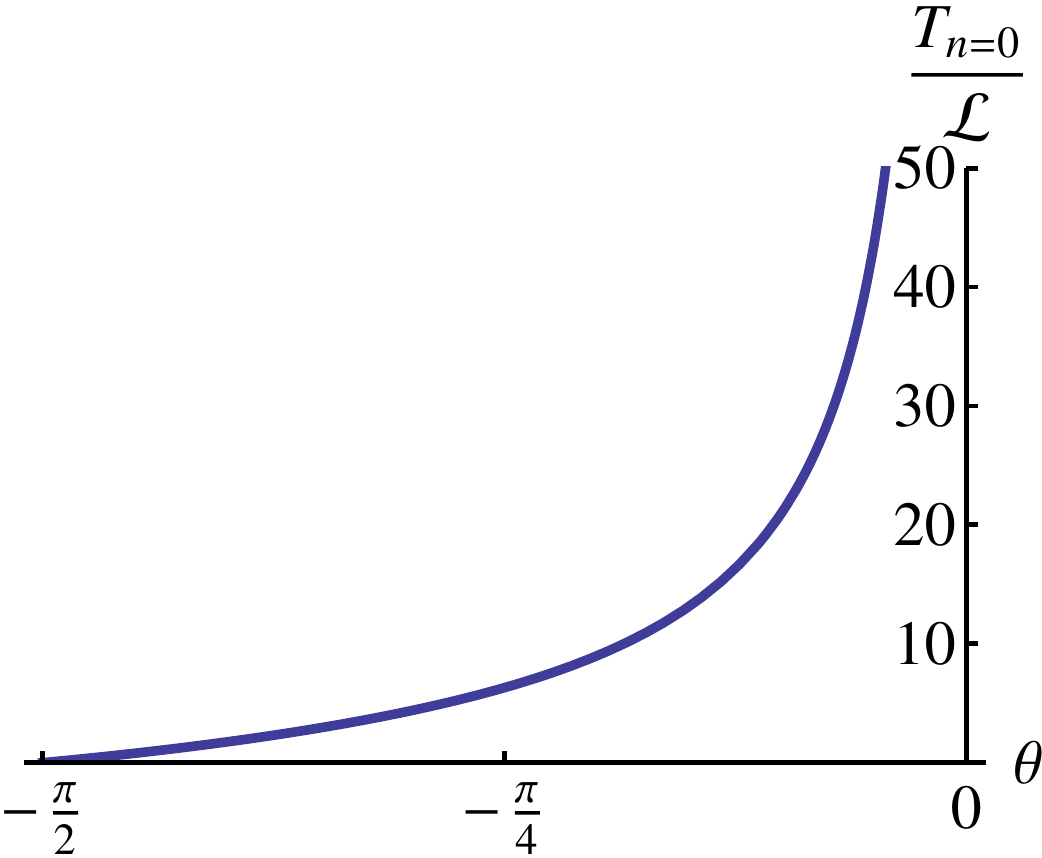}&
    \includegraphics[width=50mm]{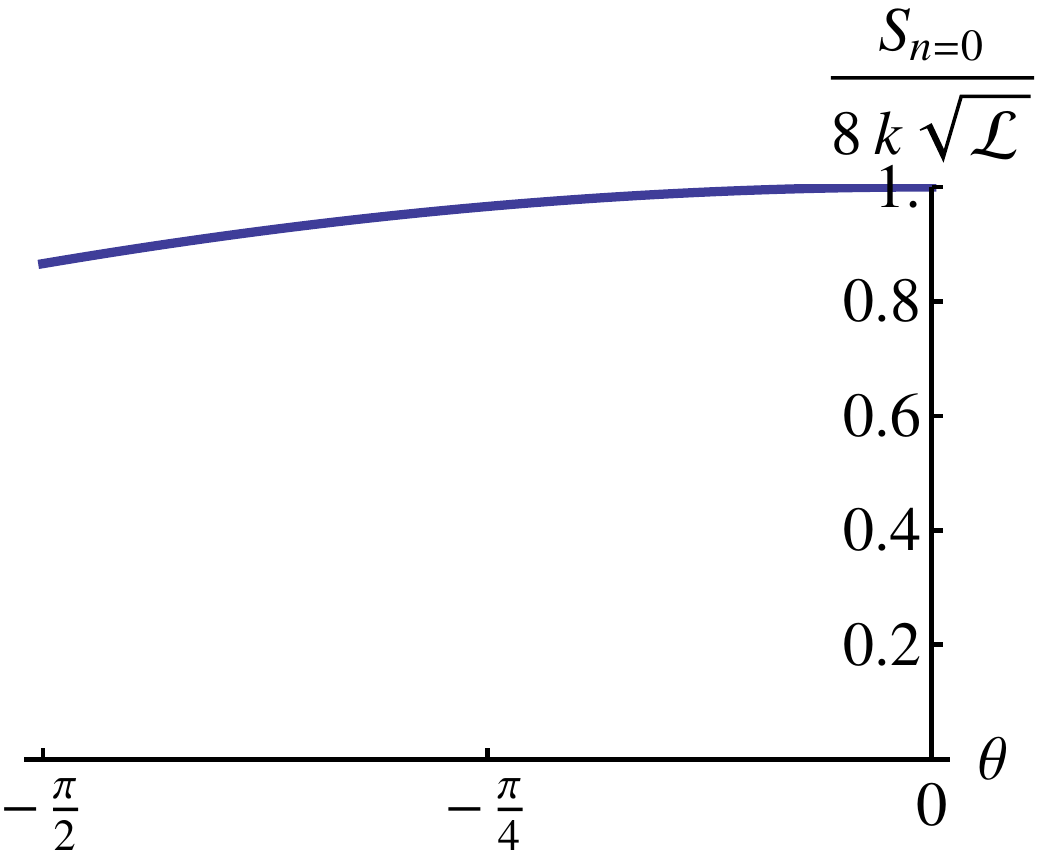}&
    \includegraphics[width=50mm]{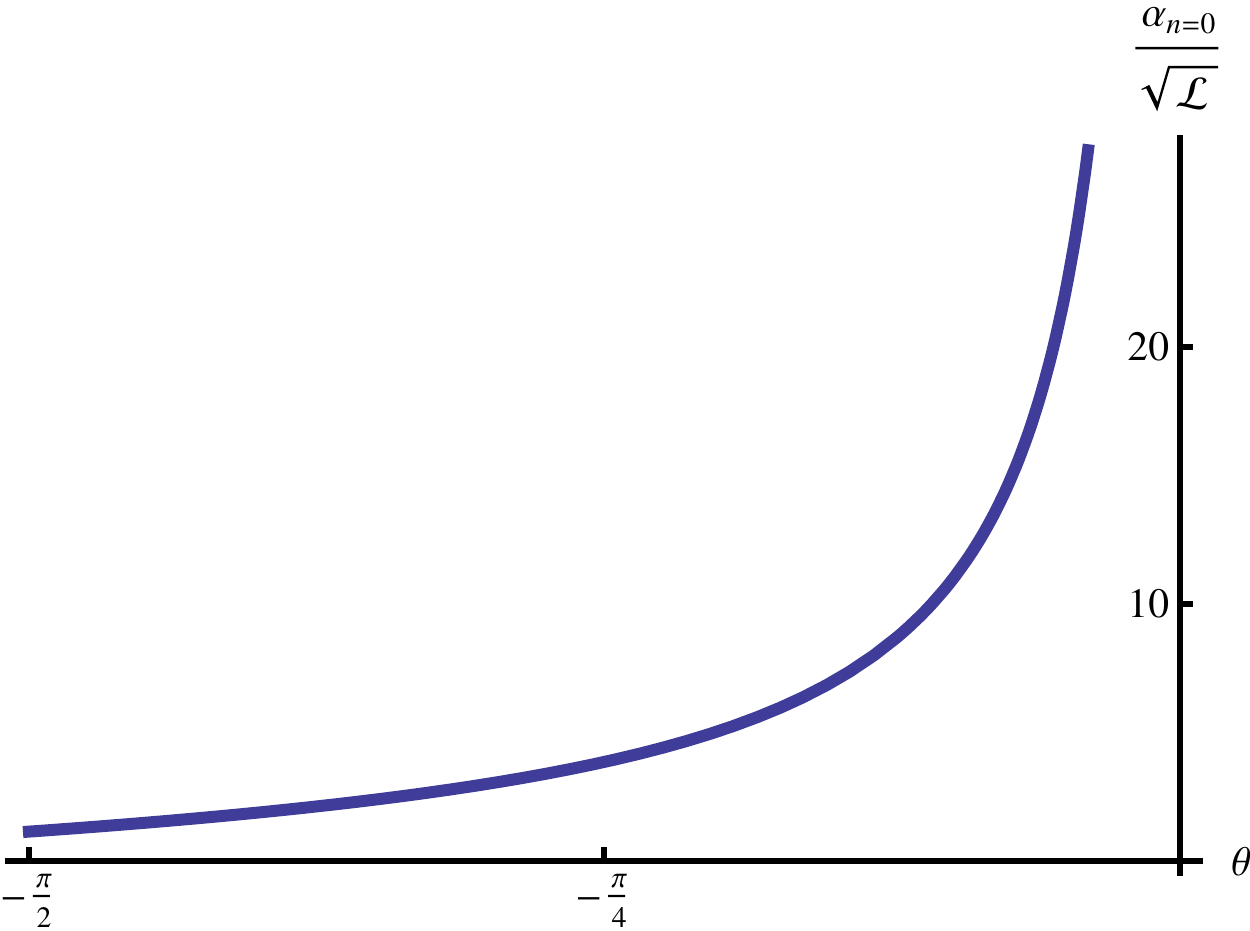}
  \end{tabular}
  \caption{Temperature, entropy and chemical potential of the $n=0$ branch}\label{fig:figbranch1}
\end{figure}
Inserting this in the second holonomy condition \eqref{holcondsb} gives us a restriction for $C$
given by
\begin{align}
  \sin\left(3C\right)=0 ,
\end{align}
which indicates that $C=n\pi/3$ for $n=0,..,5$.  Hence there are six different solutions labelled by $n$. All of these solutions can be
regarded as
thermodynamical branches of a Lifshitz black hole. The branches $n=1,2,3,4$ all show pathologies that make them unphysical. The $n=1$ case has negative temperature for all values of $\theta, {\mathcal L} $, while $n=2$ has both negative temperature and entropy. Finally,  the $n=3,4$ branches have strictly negative entropy.

Consequently, only
the branches with
$n=0$ and $n=5$ seem to be physically sensible. The entropy and temperature of the first brach ($n=0$) read
\begin{align}\label{ST0}
  S_{n=0}&=8k\sqrt{\cal L}\cos\left(\frac{\theta}{3}\right)  ,\qquad
  T_{n=0}=-\frac{4{\cal
  L}}{\sqrt{3}}\frac{\cos\theta}{\sin\frac{\theta}{3}}  ,\qquad
\alpha_{n=0}=-2\sqrt{\frac{{\cal L}}{3}}\frac{\cos{\frac{2\theta}{3}}}{\cos{\frac{\theta}{3}}}
.
\end{align}
This implies that for the temperature to be positive, one needs $-\pi/2<\theta<0$, which
imposes the constraint $-\sqrt{16{\cal L}^3/27}<{\cal W}<0$. In this range, the
entropy has its minimum at zero temperature, in accordance with the third law of
thermodynamics. Note that under this constraint, the energy \eqref{efromphi} is negative, but bounded from
below. In section \ref{bhgauge} we will discuss a simple radial gauge for which this solution looks explicitly like a
black hole. Interestingly, this gauge only exists for this branch and $n=4$, which has exactly the same entropy
but with the opposite sign. This does not mean that other branches do not have black hole gauges, as we have
not explored non-radial gauges. For now, the plots of the temperature and entropy as a function of $\theta$
for a fixed value of ${\mathcal L}$, are shown in figure \ref{fig:figbranch1}.

\begin{figure}
  \centering
  \begin{tabular}{ccc}
  \includegraphics[width=50mm]{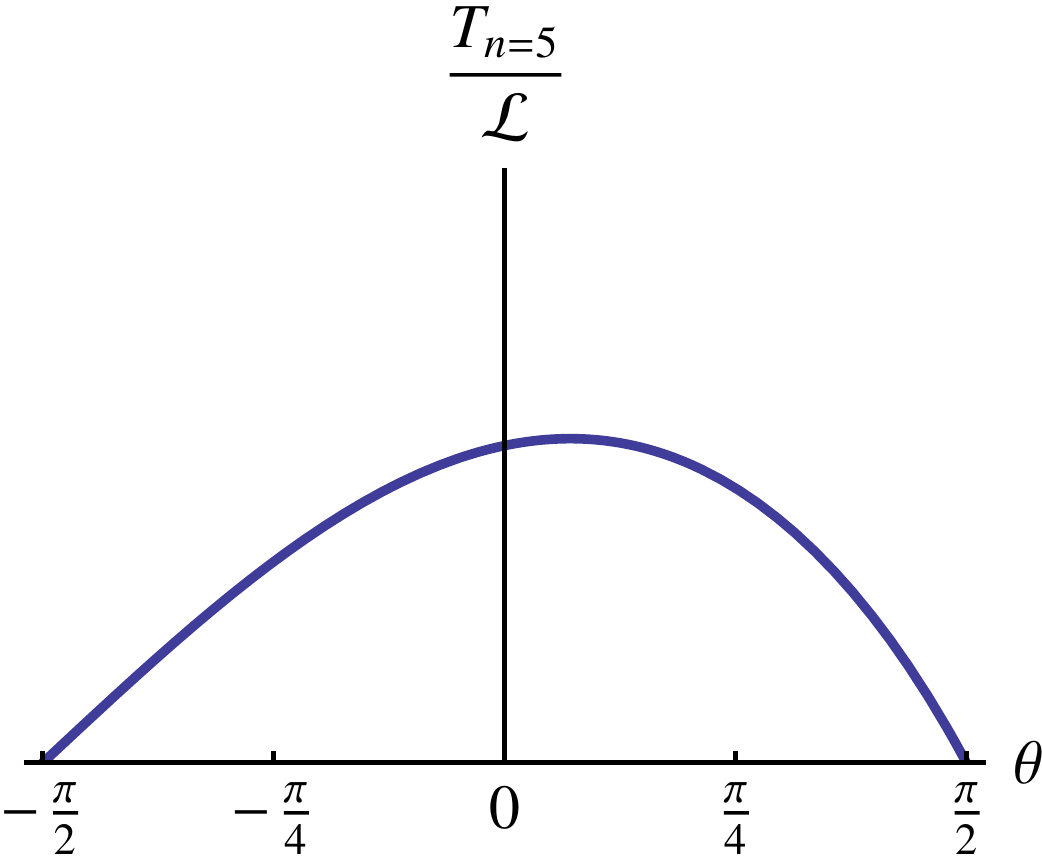}&
    \includegraphics[width=50mm]{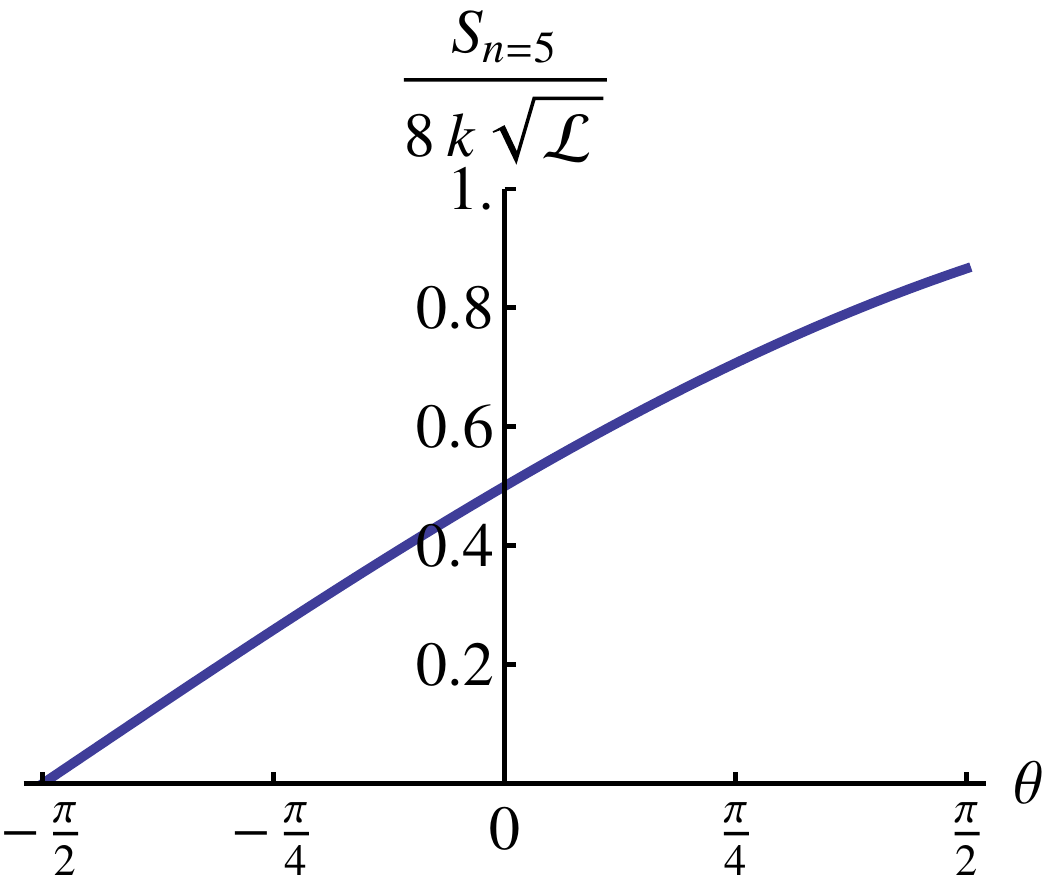}&
    \includegraphics[width=50mm]{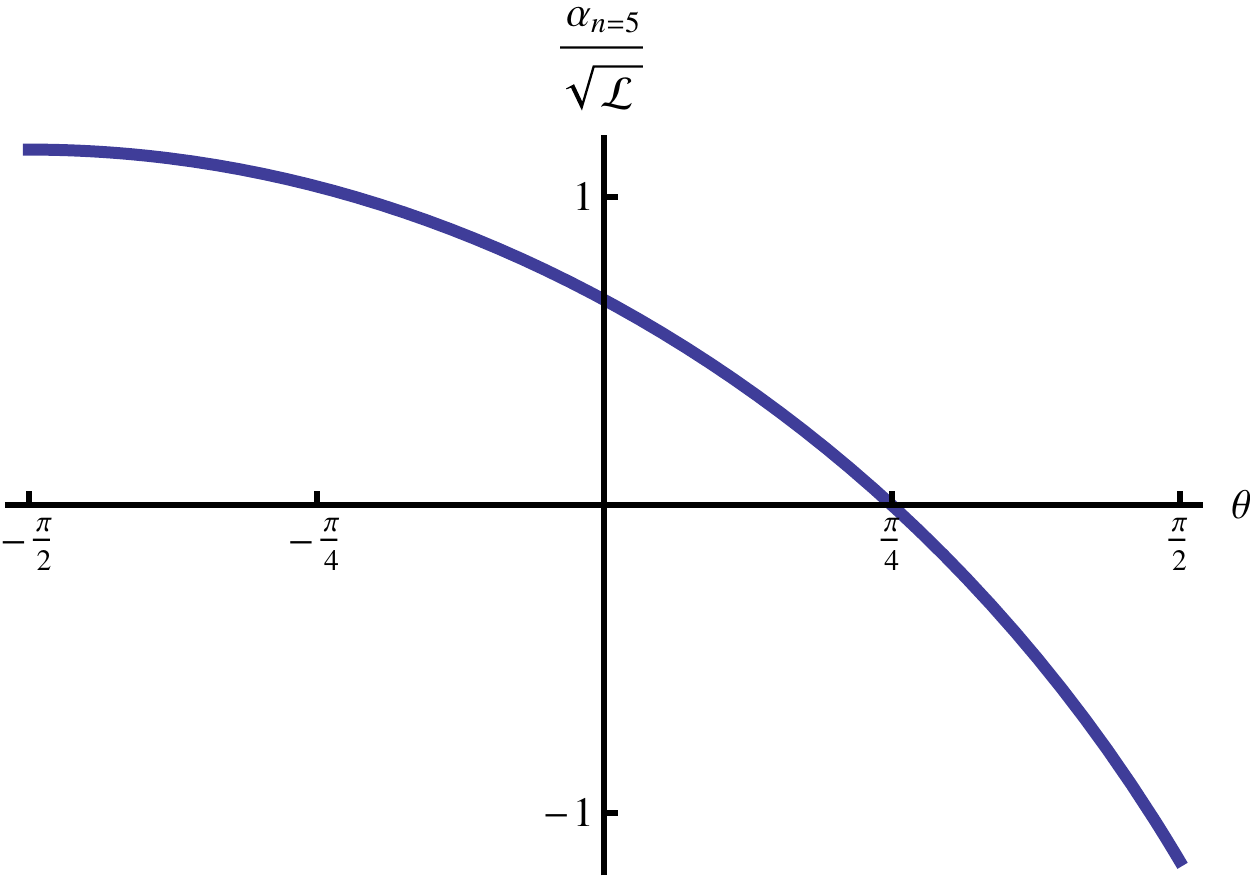}
  \end{tabular}
  \caption{Temperature, entropy and chemical potential of the $n=5$ branch}\label{fig:figbranch6}
\end{figure}

The sixth branch ($n=5$) shows the following behavior with respect to ${\cal L}$ and $\theta$
\begin{align}\label{ST5}
  S_{n=5}&=8k\sqrt{\cal L}\cos\left(\frac{\theta+5\pi}{3}\right)  ,\qquad
  T_{n=5}=\frac{4{\cal
  L}}{\sqrt{3}}\frac{\cos\theta}{\cos\frac{2\theta+\pi}{6}}  ,\qquad
\alpha_{n=5}=2\sqrt{\frac{{\cal L}}{3}}\frac{\cos{\frac{2\theta+\pi}{3}}}{\cos{\frac{2\theta+\pi}{6}}}.
\end{align}
This branch has positive values of temperature and entropy for all values of $\theta$, as shown in figure \ref{fig:figbranch6}.

\subsection{Entropy as a function of intensive parameters}\label{sec:STA}
\begin{figure}
  \centering
 \begin{subfigure}[b]{0.3\textwidth}
 \includegraphics[width=60mm]{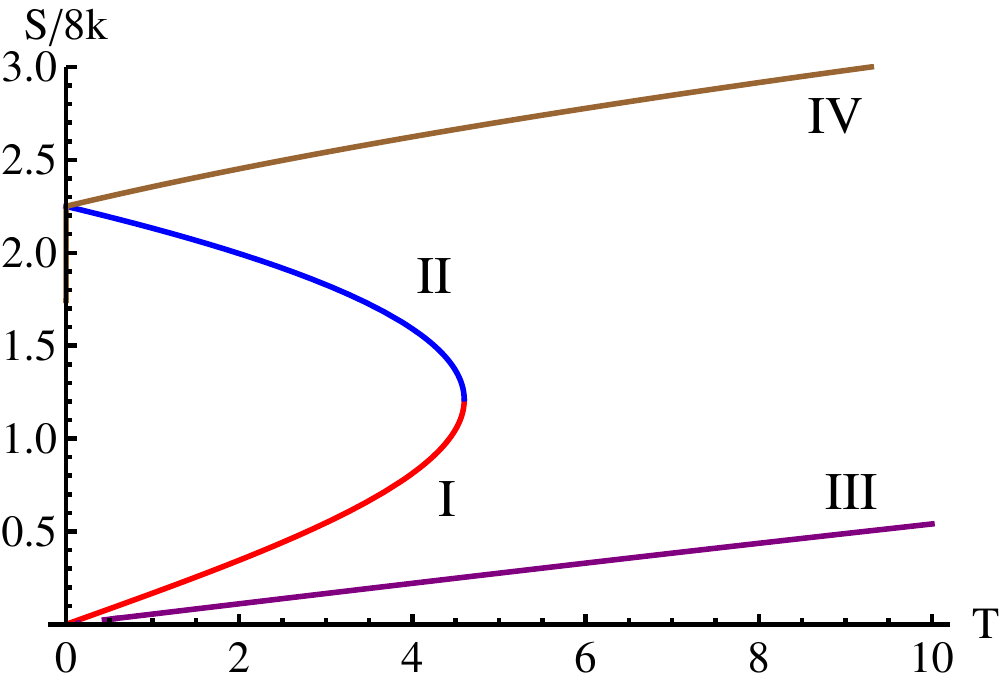}\caption{Etropies.}
\end{subfigure}\quad\quad\quad\quad\quad
 \begin{subfigure}[b]{0.3\textwidth}
\includegraphics[width=60mm]{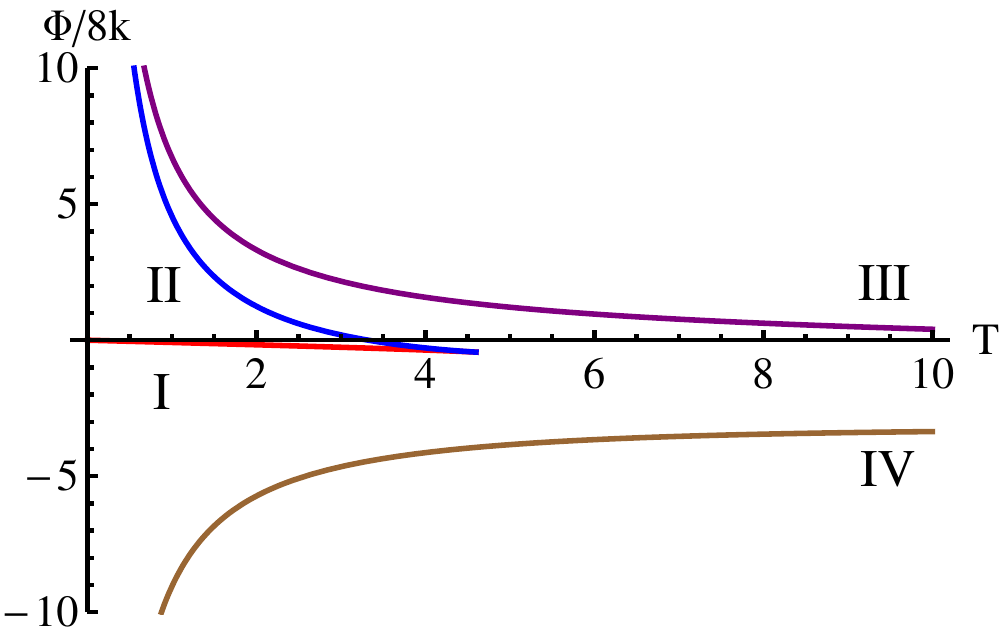}\caption{Grand potentials.}\label{subfig:potentials}
\end{subfigure}\\
 \begin{subfigure}[b]{0.3\textwidth}
  \includegraphics[width=60mm]{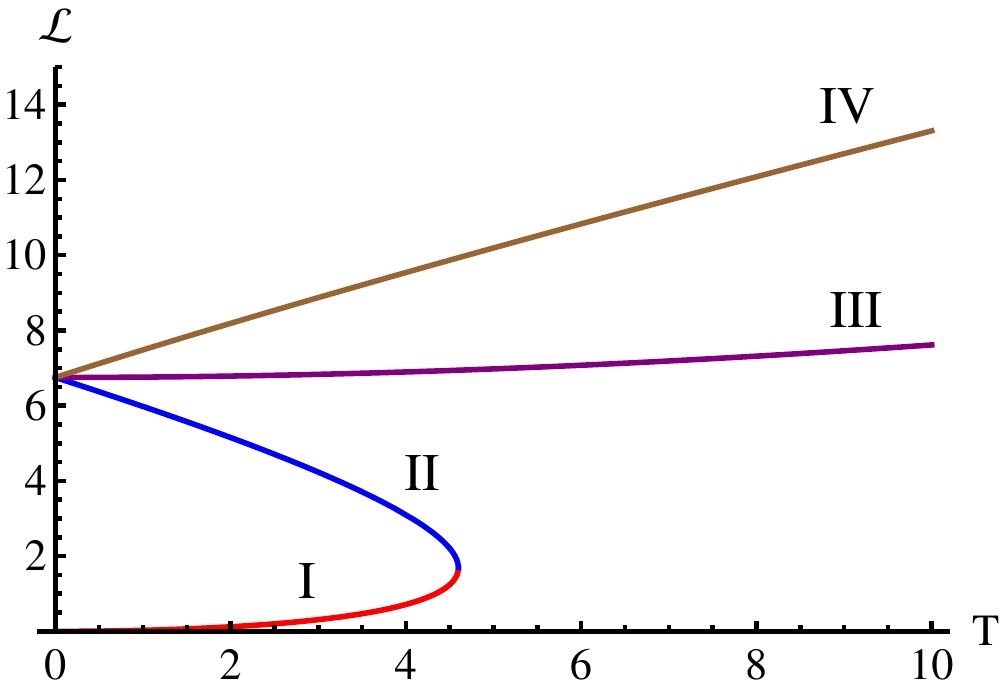}\caption{${\cal L}$ charge.}
\end{subfigure}\quad\quad\quad\quad\quad
 \begin{subfigure}[b]{0.3\textwidth}
    \includegraphics[width=60mm]{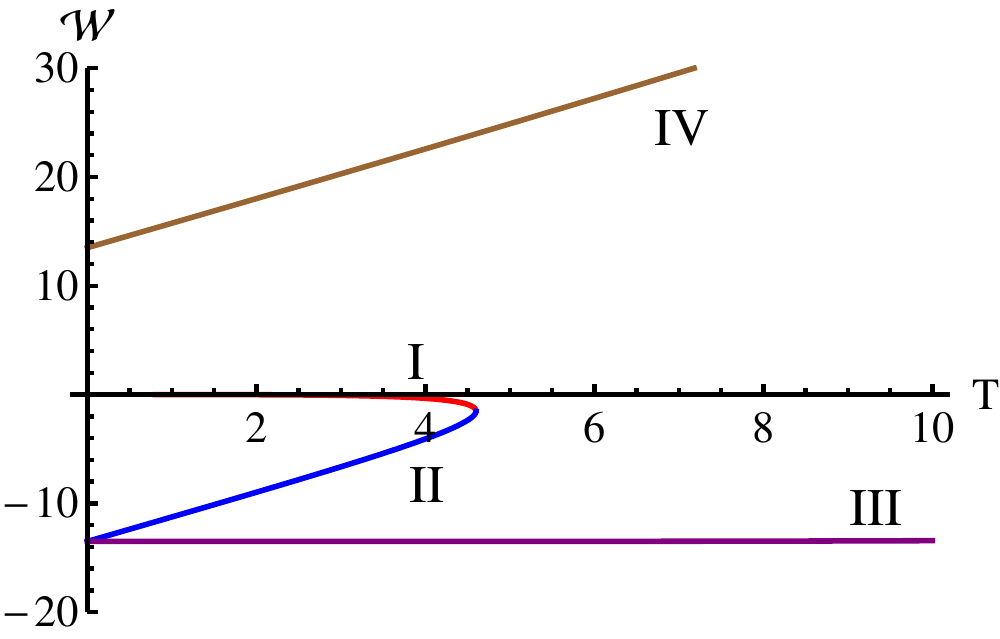}\caption{${\cal W}$ charge.}
\end{subfigure}
\caption{Entropies, Grand potentials and extensive variables for the four branches as a function of the temperature, at fixed $\alpha$. I,II and III branches have been plotted for $\alpha=3$, while branch IV has $\alpha=-3$}\label{fig:intensiveplots}
\end{figure}

Study of the stability and thermodynamical dominance of the different branches requires an expression for the entropy as a function of intensive parameters. This, in turn, requires us to solve the holonomy conditions for ${\cal L, W}
$ in terms of $\alpha, T$, and then write the entropy using \eqref{entropythermo} as a function of $\alpha$ and $T$
only. The first holonomy condition \eqref{holconds} is linear in ${\cal W}$ and can be easily solved;
\begin{align}
  {\mathcal W}= -{12 \alpha^2 {\mathcal L} + 12 {\mathcal L}^2 -3 T^2\over 36 \alpha}.
\end{align}
Plugging this into the second holonomy condition \eqref{holcondsb}, twe obtain the following quartic  equation for ${\cal L}$.
\begin{align}
  256 {\mathcal L}^4-576\alpha^2 {\mathcal L}^3+ (432 \alpha^4 -96\, T^2 {\mathcal L}^2 +(36 \alpha^2 T^2-108 \,\alpha^6){\mathcal L}+ 27 \alpha^4 T^2+ 9T^4=0.
\end{align}
This implies
the existence of four branches. Even though the number of branches is different from the ones found in last
section, one can see that appropriately gluing together these branches, one obtains the solutions we studied in section
\ref{sec:branches}. For positive temperature, the only branches with positive entropy can be found in figure
\ref{fig:intensiveplots}. Note that branch IV has been plotted for a negative value of $\alpha$ because its entropy is negative otherwise.

One can check that I and II branches map back to the $n=0$ branch from the previous section, while III and IV are related to the $n=5$ branch.
Figure \ref{subfig:potentials} shows the grand potential \eqref{grandpotential} as a function of the temperature for fixed chemical potential. In the case of negative $\alpha$, the only sensible branch is IV, and it dominates the thermodynamics. However, for a positive value of $\alpha$,  branch I ($n=0$) takes over.

We should note that the phase diagrams displayed in section \ref{sec:branches} and \ref{sec:STA} look very similar to the ones obtained for the asymptotically AdS higher spin black holes discussed in \cite{Castro:2011fm,David:2012iu,Chen:2012pc,Chen:2012ba}. This is no surprise since the holonomy equations are identical. The Lifshitz black hole  differs from the AdS  higher spin black hole however in the identification of temperature and chemical potential as well as the charges. Hence the physical interpretation of the quantities and physical reasonableness constraints (such as positive temperature) are different.

It is interesting to study the high temperature limit of these solutions. Branch I cannot reach high temperatures at fixed $\alpha$. However, in the high $\alpha$ limit, the temperature can be arbitrarily high at the
point of maximum entropy. This point is defined by a concrete value of $\theta$, so the high $\alpha$ limit can
only be reached by taking ${\cal L}$ to infinity, as can be seen by looking at equations \eqref{ST0}. In that case the
temperature grows like ${\cal L}$ while the entropy grows like $\sqrt{{\cal L}}$. This implies
\begin{align}\label{scalinga}
  S_{n=0}\sim \sqrt{T} .
\end{align}
 The same can be checked for branch IV. In the limit of high temperature, one finds that
\begin{align}
 {\cal L}&=\frac{\sqrt{3}T}{4}+3^{3/4}\sqrt{\frac{T}{32}}|\alpha|+{\cal O}(1) ,\\
  S&=-\frac{3^{1/4}\mathrm{sgn}(\alpha)}{\sqrt{8}}\sqrt{T} +{\cal O}(1).
\end{align}
Hence  for negative $\alpha$ we obtain again
\begin{align}\label{scalingb}
  S_{n=5}\sim \sqrt{T} .
\end{align}
This temperature scaling \eqref{scalinga} and \eqref{scalingb} is expected for a theory dual to a quantum field theory with $z=2$ anisotropic  Lifshitz scaling symmetry in two dimensions \cite{Gonzalez:2011nz}.

\subsection{Local stability in the grand canonical ensemble}\label{localstability}

Local thermodynamical stability is associated with the subadditivity of the entropy, as discussed
 in \cite{Gubser:2000mm,Monteiro2010a} this condition is equivalent to demanding  that the Hessian matrices of $-S$ and $-\beta\Phi$ are positive definite.
\begin{align}
  H_{mn} = \frac{\partial^2 (-S)}{\partial x_m\partial x_n}, \qquad W_{mn} = \frac{\partial^2 (-\beta\Phi)}{\partial y_m\partial y_n}
\end{align}
Which Matrix one has to consider  depends on whether one describes the thermodynamic state of the system in terms of extensive parameters $x_i$ or intensive parameters $y_i$ respectively.

In the case of our Lifshitz black hole solution, the extensive parameters can be regarded as the charges $\mathcal L$ and $\mathcal W$, while the intensive parameters can be regarded as $\beta$ and $\beta \alpha$.  Evaluation of the eigenvalues of the Hessian $H_{nm}$ for the $n=5$ branch shows that this condition can't be satisfied for any value of ${\cal L}$ and ${\cal W}$, so the $n=5$ branch is locally unstable. Demanding positive definiteness of the Hessian for the $n=0$ branch requires that $\theta \in (-3\cos^{-1}\left(\frac{3^{1/4}}{\sqrt{2}}\right),0)$. This is exactly the regime of $\theta$ covered by the curve representing branch I in figure \ref{fig:intensiveplots}.

One can further check that this result is consistent with the description in terms of potentials. Computation of the eigenvalues of $W_{nm}$ for the four branches studied in section \ref{sec:STA}, indeed shows that branch I is locally stable, while II is not.

\subsection{Metric and black hole gauge}\label{bhgauge}
We now investigate the question whether  a gauge exists  in which the metric of  the Lifshitz black hole
solutions  displays a regular horizon. In fact, we demonstrate that for some  branches  one can maintain
radial gauge and
choose some of the residual gauge such that  $g_{tt}$ contains a double zero and $g_{xx}$ is
regular.

We begin again with the ansatz \eqref{alant}, \eqref{alanx} and the flatness
conditions  \eqref{flat},
where again the barred sector is determined by the non-rotating condition $\bar
a_x = -a_x^T$ and
$\bar
a_t=a_t^T$.  We also regard equations \eqref{gaugeinva} as a reparametrization of $w_{t,2},l_{t,1},l_{x,-1},$ and $w_{x,-2}$ as functions of the charges and potentials ${\cal L},{\cal W},\mu_1,\mu_2,$ and the residual gauge parameter $w_{x,0}$. Next, we solve for the value of $w_{x,0}$ for which the corresponding
metric derived
from
\eqref{metform} has a double zero in $g_{tt}$ at some value of $\rho=\rho_h$, the
location of the
corresponding horizon.  To do this, first we note that the metric component
$g_{tt}$ can be written
as
\begin{align}
  g_{tt} = -(e^{2\rho} p_1 - e^{-2\rho}p_2)^2 - (e^\rho p_3 - e^{-\rho} p_4)^2
  \label{bhgtt}
\end{align}
where $p_i$ are $\rho$-independent coefficients given by
\begin{align}
  p_{1} &=\mu _{2} \\
  p_{3} &=\mu _{1}-\mu _{2}\frac{w_{x,0}}{3} \\
  p_{2} &=-\left( \frac{w_{x,0}}{3}\right) ^{3}\frac{\mu _{1}}{4}+\mathcal{W%
  }\mu _{1}+\mathcal{L}^{2}\mu _{2}+\left( \frac{w_{x,0}}{3}\right) ^{4}%
  \frac{\mu _{2}}{16}+\frac{w_{x,0}}{3}2\mathcal{W}\mu _{2}+\frac{1}{2}%
  \mathcal{L}\frac{w_{x,0}}{3}\left( 2\mu _{1}+\frac{w_{x,0}}{3}\mu
  _{2}\right) \\
  p_{4} &=\mathcal{L}\mu _{1}-\left( \frac{w_{x,0}}{3}\right) ^{2}\frac{3}{4%
  }\mu _{1}+\frac{w_{x,0}}{3}\mathcal{L}\mu _{2}+\left( \frac{w_{x,0}}{3}%
  \right) ^{3}\frac{\mu _{2}}{4}+2\mathcal{W}\mu _{2}
\end{align}
It is clear that $g_{tt}$ is zero if and only if each term in parentheses on the
right hand side of
\eqref{bhgtt} is zero for the same value of $\rho_h$ which implies that $p_2/p_1 =
(p_4/p_3)^2$.
Using
the expressions above for $p_1, \dots, p_4$, this constraint is equivalent to the
following cubic
equation for $w_{x,0}$:
\begin{align}
  w_{x,0}^3 - 36\mathcal L w_{x,0} - 108\mathcal W = 0.
\end{align}
The three solutions are given by
\begin{align}\label{w0xsols}
w_{x,0}=4\sqrt{3{\cal L}}\cos\left(\frac{\cos^{-1}(\sin\theta)}{3}+m \frac{
2\pi}{3}\right),
\end{align}
with $m=0,1,2$. However the only solution with a positive and real horizon
$\rho_h=\sqrt[4]{p_2/p_1}=\sqrt{p_4/p_3}$ is the one with $m=2$, which can be
simplified to
\begin{align}\label{w0xsolm2}
w_{x,0}=-4\sqrt{3{\cal L}}\sin\left(\frac{\theta}{3}\right).
\end{align}

The horizon is then located at
\begin{align}\label{rhoh}
\rho_h=\sqrt{{\cal L}\left(2\cos\frac{2\theta}{3}-1\right)}.
\end{align}
It seems that we did not need to impose the holonomy conditions in order to find
this black hole gauge. However, we still need to check that the metric and the
spin three field in this gauge are smooth around the cycle $t\sim t+2\pi i$. this
implies the following conditions
\begin{align}
1=\sqrt{\frac{g_{tt}}{-2g_{\rho\rho}}}\bigg|_{\rho_h}  ,\quad \quad
1=\sqrt{\frac{\psi_{xtt}}{-2\psi_{x\rho\rho}}}\bigg|_{\rho_h}.
\end{align}
Direct substitution of the charges and sources for the six branches found in
previous sections shows that only the $n=0,3$ cases satisfy these identities. This
can mean that this gauge is appropriate for those two solutions, while the other
branches require giving up the radial gauge chosen in equation \eqref{bigadef}. As
we have argued in section \ref{sec:branches}, the $n=3$ branch does not seem to be
physically sensible. For this reason we will focus our attention in branch $n=0$.
The values of the spin fields at the horizon in this branch obey the following
relations
\begin{align}
g_{tt}|_{\rho _{h}} &=0 , \qquad
g_{tt}^{\prime }|_{\rho _{h}} =0 , \qquad
g_{xx}|_{\rho _{h}} =4\mathcal{L} , \qquad
\psi _{xxx }|_{\rho _{h}} =2\mathcal{W}.
\end{align}
So we can recast our expresion for the entropy as
\begin{align}\label{entropyarea}
  S=\frac{4k}{\pi}A\cos\left[\frac{1}{3}\sin^{-1}\left(\frac{3^{\frac{3}{2}}\psi_3}{A^3}\right)\right]
\end{align}
where
\begin{align}
  A&=2\pi\sqrt{g_{xx}|_{\rho _{h}}},\qquad
  \psi_3=\psi _{xxx }|_{\rho _{h}},
\end{align}
which is very similar to the entropy formula found for asymptotically AdS higher spin black holes \cite{Perez:2013a}. It would be interesting to investigate whether the local thermodynamic instability of the $n=5$ branch discussed in   section \ref{localstability} and the absence of a regular horizon are related. However, it is an open and interesting question, if  for the $n=5$ branch there is a more general radial gauge choice (along the lines of \cite{Ammon:2011nk}) which has  a regular horizon.

\section{Generalizations}
\setcounter{equation}{0}
\label{sec5}
In this section we will present some observations on possible generalizations of
our $SL(3,\reals)$
results obtained in the previous sections.

\subsection{Rotating solutions}
In the present paper we have limited ourselves to  non-rotating solutions, for
which the connections
$A$ and $\bar A$ are related by equation \eqref{conrel}. Since the two Chern-Simons
connections $A,\bar
A$
are independent, it is clear that constructing a solution with angular momentum
entails lifting the
condition \eqref{conrel}.  This also means that there will be two holonomy
conditions
\label{trivhol}
for the $A$ and the $\bar A$ connection.
Recall that in the  $SL(3,\reals)\times SL(3,\reals)$ black hole first discussed
in
\cite{Gutperle:2011kf} a rotating higher spin black hole is obtained by
choosing  modular parameter to be complex  $\tau = \Omega+ i\beta$, where $\Omega$
is the potential
dual to the angular momentum. For the Lifshitz black holes this cannot work quite
the same way and
we
present some observations  here. Note that in the holographic dictionary or the
stress energy
complex
of a Lifshitz theory \eqref{trivhol} the angular momentum (i.e. the momentum along
the $x$ direction if
we
take $x$ to be compact) is identified with ${\mathcal L}- \bar {\mathcal L}$,
whose conjugate
potential
is $\mu_1 -\bar \mu_1$ and the energy is  identified with ${\mathcal W}+ \bar
{\mathcal W}$, whose
conjugate potential is $\mu_2+\bar \mu_2$. Hence it is likely that a rotating
solution can be
constructed by choosing a connections with $\mu_1\neq \bar \mu_1$ and keeping the
indentification of
the temperature $\beta$ the same as in the non-rotating case. The expressions for
the metric and
higher
spin fields are  much more complicated.  This implies also that the analysis of
the black hole gauge
done section \ref{bhgauge} becomes more involved, and we leave these questions
for future work.
We
also note that, to our knowledge,  no rotating Lifshitz black hole solutions have
been constructed
using the standard supergravity actions. Hence constructing such solutions in
higher spin gravity
might
be interesting.

\subsection{Lifshitz vacuum for $hs(\lambda)$}

In this section we discuss some steps in generalizing the construction of Lifshitz
black holes from
$SL(3,\reals)$ to $hs(\lambda)$, note that this generalization will also include
the case of
$SL(N,\reals)$ by choosing $\lambda=N$, where the infinite-dimensional Lie algebra
reduces to
$SL(N,\reals)$. Our conventions for $hs(\lambda)$ are summarized in appendix
\ref{hslamconv}.

A Lifshitz vacuum in the $hs(\lambda)$ theory can be easily constructed as
follows
\begin{align}\label{hsconn}
a_t &= {1\over \sqrt{\tr(V^s_{s-1} V^s_{-(s-1)})}} V^s_{s-1} ,   &a_x&=
{1\over
\sqrt{\tr(V^2_{1} V^2_{-1})}}  V_1^2\nonumber\\
\bar a_t &= {1\over \sqrt{\tr(V^s_{s-1} V^s_{-(s-1)})}} V^s_{-(s-1)} ,
&\bar a_x&=
-{1\over
\sqrt{\tr(V^2_{1} V^2_{-1})}}  V_{-1}^2.
\end{align}
Note that since
\begin{align}
  [V_1^2, V^s_{s-1}]_\star=0, \quad \quad [V_{-1}^2, V^s_{-(s-1)}]_\star=0,
\end{align}
this satisfies the flatness condition for a connection in the radial gauge.  The
gauge connections
$A_\mu$ and the metric are obtained from \eqref{hsconn}  by adapting the formulae
\eqref{metform}
and
using   $b= \exp( \rho V^2_0)$ It follows that the metric is of the form.
\begin{align}
  ds^2 = -e^{2 (s-1) \rho} dt^2+ e^{2\rho}dx^2+ d\rho^2.
\end{align}
Hence we can realize an asymptotically Lifshitz metric in the $hs(\lambda)$ theory
for any
$z=2,3,4,\cdots$, by setting $s=z+1$.
Note that some higher spin fields will be non-vanishing for this $hs(\lambda)$
Lifshitz vacuum. By
setting $\lambda=N$, the infinite-dimensional $hs(\lambda)$ gauge algebra
truncates to a finite-dimensional $SL(N,\reals)$, and the connections give Lifshitz vacua with
$z=N-1,N-2,\cdots,2$.

\subsection{An $hs(\lambda)$ Lifshitz black hole}
Here we limit ourself to the BH for $z=2$, which is related to the $hs(\lambda)$
black hole with a
chemical potential for the spin three charge, which is most extensively studied
in the literature.
The connection is given by
\begin{align}\label{hsconnb}
 a_x &= V_1^2+  \tilde {\mathcal L} V_{-1}^2+ \tilde{\mathcal W} V_{-2}^3+ \tilde
 {\mathcal U}
 V_{-3}^4+\cdots\\
 a_t &=  \tilde \mu_1 a_x+
\tilde \mu_2 (a_x \star a_x)\mid_{\rm traceless}.
\end{align}
Here, $ \tilde{\mathcal L} , \tilde{\mathcal W}, \tilde{\mathcal U}$, etc are
associated with
charges
of spin $2,3,4, \cdots$. We have tilded all quantities to distinguish them from
the quantities
appearing in the higher spin black hole reviewed in the appendix
\ref{hslamconv}.

By construction the connection \eqref{hsconnb} satisfied the flatness condition.
To define a regular
black hole in a higher spin Chern-Simons theory one has to impose a holonomy
condition on the gauge
connection around the euclidean time circle.
The holonomy condition which we choose is again that the holonomy  is equal to the
BTZ holonomy for
the
$hs(\lambda)$ black hole defined in appendix \ref{hslamconv}. One might object
that in the case of
the
Lifshitz BH this condition  seems  less well motivated since there is no analog of
a BTZ black hole
for
an asymptotically  Lifshitz spacetime, however a better way to think about this
is that the BTZ
holonomy simply states that the holonomy of the BH is in the center of
$hs(\lambda)$ (see
{Gaberdiel:2013jca}  for a discussion on how the center of $hs(\lambda)$ is
defined).

If we compare the holonomy associated with $a_t$  defined in \eqref{hsconnb} and the
higher spin black
hole
holonomy \eqref{holon} one recognizes that they are the same upon the following
identifications
\begin{align}
  \tilde \mu_1= 2\pi \tau ,\quad \quad \tilde \mu_2 = -2\pi \alpha.
\end{align}
Furthermore the charges can also be identified
\begin{align}
  \tilde {\mathcal L} = -{2\pi \over k} {\mathcal L} , \quad \quad  \tilde {\mathcal W} =-{\pi \over 2k} {\mathcal W} , \quad \cdots
\end{align}
Since there is a one-to-one map of parameters one might ask how this can be
different than the
$hs(\lambda)$   \cite{Kraus:2011ds}.
The answer lies in the fact that while (this was true for the $SL(3,\reals)$
case too) the
holonomy
conditions have the same functional form, the interpretations of $\tilde\mu_1$ and
$\tilde \mu_2$ are
different. The inverse temperature $\beta$ and the  chemical
potential $\tilde \alpha$ can be related to $\tilde \mu_1$ and $\tilde \mu_2$ following the 
 the $SL(3,\reals)$ Lifshitz black hole example
\begin{align}
  \tilde \mu_1=  \beta \tilde \alpha   ,\quad \quad \tilde \mu_2 =  \beta.
\end{align}
This means that the most natural regime for the Lifshitz black hole , i.e.
$\tilde\beta$ finite and
$\tilde \alpha$ small, is not  the same regime  as the one which allows the
perturbative solution
of
the holonomy conditions first obtained in \cite{Kraus:2011ds}. Indeed if we take
the limit $\tilde
\alpha \to 0$, this is equivalent for the higher spin black hole to taking the limit $
\tau \to 0$ and
keeping
$\alpha$ finite, i.e. taking an infinite temperature limit and finite chemical
potential.

\section{Discussion}
\setcounter{equation}{0}
\label{sec6}

In this paper we have discussed the construction of holographic spacetimes dual
to field theory with
Lifshitz $z=2$
scaling symmetry . In addition we have constructed black hole solutions in these theories.
One interesting
feature
of these theories is that the connections, holonomy conditions and thermodynamic
relations are all
very
similar to the higher spin black holes first constructed in
\cite{Gutperle:2011kf}. This can be
traced
back
to the fact that the Lifshitz black hole connections and the higher spin black hole
connections are related
by
replacing $t,x$ by $\bar z,z $ respectively. Note however that the interpretation
of the parameters
is
quite different. First, the holographic identification of the stress energy
complex of the QFT with
Lifshitz
symmetry and the role of the fields ${\mathcal L}$ and ${\mathcal W}$ are quite
different for the
Lifshitz
theory compared to the $W_3$ CFT. Second, for the Lifshitz black hole solutions
the identification
of
the temperature and higher spin chemical potential is in some sense reversed
compared to the higher
spin black hole, this leads to a different interpretation of the thermodynamics.
The solution of the holonomy conditions has different branches, which we can interpret as different thermodynamic phases. We have shown that only one branch (branch I of section \ref{sec:STA}) has  1. positive entropy and 2. positive  temperature, 3. is locally thermodynamically stable and 4. enjoys a radial gauge with a regular horizon. All other branches do not satisfy one or more of these conditions and are therefore physically not satisfying.

We have briefly discussed generalizations of the black hole solutions found in
this paper.
It would be interesting to study  Lifshitz black hole solutions  in $hs(\lambda)$
further, since
there
exists  a concrete proposal for a dual CFT and the Lifshitz theories could be
interpreted as
deformations of the  CFT. Furthermore since it is possible to couple scalar matter
consistently there
are
independent probes of the geometry of the black hole. To make progress one has to
solve the holonomy
conditions either exactly or maybe less ambitiously determine wether it is
possible to solve the
holonomy conditions perturbatively for small $\tilde \alpha$ and finite
temperature. We plan to
return
to these interesting  questions in the future.

\bigskip

\noindent{\Large \bf Acknowledgements}

\bigskip

This  work was in part supported  by NSF grant PHY-07-57702 and PHY-13-13986. M.G. is grateful to
the
Centro de Ciencias de Benasque Pedro Pascual for hospitality while this work was
in progress.  M.G.
is
gratefu for hospitality at the Institute
of Theoretical
Physics, University of Jena while this paper was finalized.  E.H. acknowledges
support from
Fundaci\'on
la Caixa. We are grateful to Martin Ammon, Per Kraus, Edgar Shaghoulian, and Arnaud Lepage-Jutier for useful conversations.
\bigskip

\newpage

\appendix

\section{Conventions}
\setcounter{equation}{0}
\label{appendixa}
In this appendix we present some details on the conventions and explicit
representations of the Lie
algebras used in the main body of the paper.

\subsection{Explicit $SL(3,\reals)$ representation}

The $SL(2,\reals)$  generators of the principal embedding  are given by the
following matrices
\begin{align}
  L_{-1} = \begin{pmatrix}
          0 & \sqrt 2 & 0 \\
          0 & 0 & \sqrt 2 \\
          0 & 0 & 0 \\
        \end{pmatrix}, \qquad
  L_1 = \begin{pmatrix}
          0 & 0 & 0 \\
          -\sqrt 2 & 0 & 0 \\
          0 & -\sqrt 2 & 0 \\
        \end{pmatrix}, \qquad
  L_0 = \begin{pmatrix}
          1 & 0 & 0 \\
          0  & 0 & 0 \\
          0 & 0  & -1 \\
        \end{pmatrix}.
\end{align}
and the spin 3 generators, on which we omit the superscript $^{(3)}$ for
notational simplicity, are
as
follows:
\begin{align}
  W_{-2} &= \begin{pmatrix}
          0 & 0 & 2 \\
          0  & 0 & 0 \\
          0 & 0  & 0 \\
        \end{pmatrix}, \qquad
  W_{-1} = \begin{pmatrix}
          0 & \frac{1}{\sqrt 2} & 0 \\
          0 & 0 & -\frac{1}{\sqrt 2}\\
          0 & 0 & 0 \\
        \end{pmatrix}, \qquad
  W_0 = \begin{pmatrix}
          \frac{1}{3} & 0 & 0 \\
          0  & -\frac{2}{3} & 0 \\
          0 & 0  & \frac{1}{3} \\
        \end{pmatrix} \\
  W_1 &= \begin{pmatrix}
          0 & 0 & 0 \\
          -\frac{1}{\sqrt 2} & 0 & 0 \\
          0 & \frac{1}{\sqrt 2} & 0 \\
        \end{pmatrix}, \qquad
  W_2 = \begin{pmatrix}
          0 & 0 & 0 \\
          0  & 0 & 0 \\
          2 & 0  & 0 \\
        \end{pmatrix}.
\end{align}
If we define $(T_1,T_2, \dots, T_8) = (L_1, L_0, L_{-1}, W_2, \dots W_{-2})$, then
traces of all
pairs
of generators are given by
\begin{align}
  \mathrm{tr}(T_iT_j)
  &=
  \left(\begin{array}{ccc|ccccc}
        &   & -4  &   0  &    &        \cdots       &       &0\\
            & 2   &      &  \vdots  &    &    \ddots           &       &\vdots\\
    -4  &   &     &    0  &    &        \cdots       &       &0\\
       \hline
         0   & \cdots  &  0   &     &   &               &       & 4\\
            &   &     &     &   &               & -1    &\\
         \vdots   & \ddots  &  \vdots   &     &   & \frac{2}{3}   &       &\\
            &   &     &     & -1&               &       &\\
          0  & \cdots  &   0  & 4   &   &               &       &
     \end{array}\right)
\end{align}

\subsection{$hs(\lambda)$ conventions and black hole}\label{hslamconv}
 Here we follow the conventions of \cite{Kraus:2012uf} and
 \cite{Gaberdiel:2013jca}. The main
 formulas
 we use are, the  lone star products
 \begin{align}
 V_m^s \star V_n^t= {1\over 2} \sum_u{1,2,\cdots}^{s+t-|s-t|-1}
 g_u^{st}(m,n;\lambda)
 V_{m+n}^{s+t-u}.
 \end{align}
 The star product is used to define the commutator between Lie algebra generator
 and is denotes by
 $[\cdot,\cdot]_\star$.
 For the elements of the Lie-algebra $V^s_m$ one has $|m|<s$ (the generators are
 zero otherwise).
 The elements $V^2_{-1,0,1}$ form a SL(2,R) sub algebra and $V^s_m$ form spin s
 representation
 \begin{align}
 [V_m^2, V_n^t]_*=\Big( m(t-1)-n\Big) V_{m+n}^t.
 \end{align}
 The algebra has a unit element denoted by $X^1_0$, the trace is defines by
 \begin{align}
 Tr(X)=X{ \mid} {}_{V^1_0}.
 \end{align}
  A $hs(\lambda)$ black hole with a chemical potential for the spin 3 charge
  (this can be
  generalized
  to arbitrary spin $s$) has the following connections
 \begin{align}
 a_z&= V_1^2-{2\pi \over k} {\mathcal L} V_{-1}^2 -{\pi \over 2k} {\mathcal W}
 V_{-2}^3 + {\mathcal
 U}V_{-3}^4+\cdots,\nonumber\\
 a_{\bar z} &= -{\alpha\over \bar \tau} (a_z \star a_z)\mid_{\rm traceless}.
 \end{align}
 The holonomy around the time circle is given by $H=e^\omega$ with
 \begin{align} \label{holon}
   \omega=2\pi\big(\tau a_z+ \bar\tau a_{\bar z}\big).
 \end{align}
 The holonomy condition for the black hole is that the holonomy is the same as the
 holonomy of the
 BTZ
 black hole
 \begin{align}
   H=H_{BTZ}.
 \end{align}
 where $\omega_{BTZ}$ is given by
 \begin{align}
   \omega_{BTZ}=2\pi  \tau V_1^2 + {\pi\over \tau} V_{-1}^2.
 \end{align}
 This condition is equivalent to the following conditions on the powers of
 $\omega$ (see eq. 2.17 of
 \cite{Gaberdiel:2013jca}).
 \begin{align}
 Tr(\omega^n) = {1\over \lambda} \lim_{t\to 0} \left( \partial_t^n {\sin \pi
 \lambda t\over \sin \pi
 t}\right).
 \end{align}
 These conditions have been solved perturbatively in the chemical potential
 $\alpha$ and one gets
 the
 charges ${\mathcal L},{\mathcal W},{\mathcal U},\cdots$  as a power series in
 $\alpha$ (and
 depending
 on $\tau$), such that as $\alpha\to 0$ one gets back the BTZ black hole.

\end{document}